\documentclass[12pt, draftclsnofoot, onecolumn]{IEEEtran}
\usepackage{graphicx}
\usepackage{amsmath}
\usepackage{amssymb}
\usepackage{mathrsfs}
\usepackage{stfloats}
\usepackage{dsfont}
\usepackage{setspace}
\usepackage{array}
\usepackage{bbm}
\usepackage{bm}
\usepackage{multirow}
\usepackage{subfigure}
\usepackage{xcolor}

\usepackage{accents}

\usepackage[centerlast,small]{caption}
\usepackage{hyperref}

\usepackage{amsthm}

\theoremstyle{remark}

\theoremstyle{remark}

\theoremstyle{remark}

\theoremstyle{remark}

\theoremstyle{plain}
\allowdisplaybreaks[4]

\usepackage{algorithm,algorithmic}

\begin{document}
%
%
%
%
%
%
\title{K-Means Clustering-Aided Non-Coherent Detection for Molecular Communications}
\author{
Xuewen~Qian,
Marco~Di~Renzo,~\IEEEmembership{Fellow,~IEEE}, and Andrew Eckford,~\IEEEmembership{Senior~Member,~IEEE}

\thanks{Manuscript received Aug. 30, 2020; revised Jan. 19, 2021; and March 12, 2021. X. Qian and M. Di Renzo are with Universit\'e Paris-Saclay, CNRS and CentraleSup\'elec, Laboratoire des Signaux et Syst\`emes,  91192 Gif-sur-Yvette, France. (e-mail: marco.di-renzo@universite-paris-saclay.fr). A. Eckford is with the Dept. of Electrical Engineering and Computer Science, York University, 4700 Keele Street, Toronto, Canada M3J 1P3. }}

\markboth{Transactions on Communications} {X. Qian, M. Di Renzo, and A. Eckford: K-Means Clustering-Aided Non-Coherent Detection for Molecular Communications}
\maketitle
%
%
%
\begin{abstract}
In this paper, we consider non-coherent detection for molecular communication systems in the presence of inter-symbol-interference. In particular, we study non-coherent detectors based on memory-bits-based thresholds in order to achieve low bit-error-ratio (BER) transmission. The main challenge of realizing detectors based on memory-bits-based thresholds is to obtain the channel state information based only on the received signals. We tackle this issue by reformulating the thresholds through intermediate variables, which can be obtained by clustering multi-dimensional data from the received signals, and by using the K-means clustering algorithm. In addition to estimating the thresholds, we show that the transmitted bits can be retrieved from the clustered data. To reduce clustering errors, we propose iterative clustering methods from one-dimensional to multi-dimensional data, which are shown to reduce the BER.  Simulation results are presented to verify the effectiveness of the proposed methods.
\end{abstract}
%
%
%
%
%
%
%
%
\section{Introduction} \label{Introduction} 
Recent developments in biology and nano-technology have gained attention in  micro-scale communications in light of emerging potential applications, e.g., bio-robots \cite{nelson2004biological}. In this context, molecular communication (MC) is regarded as a promising solution \cite{Farsad2016} because of the difficulty of using traditional electromagnetic-based techniques \cite{Hiyama2005}. In a MC system, the information is disseminated via small particles \cite{Farsad2016}. Among the different modulation schemes that can be used in MC systems \cite{kuran2011modulation}, binary concentration shift keying (CSK) modulation is the simplest method for encoding information onto the number of released particles. In addition, diffusion via Brownian motion \cite{pierobon2013capacity} is the most common solution for allowing information particles propagate from a transmitter to a receiver.

\subsection{Motivation}
In diffusion-based MC systems, there exist several issues to be solved. A major issue is the non-negligible inter-symbol interference (ISI) that is caused by the intrinsic characteristics of channels with memory. If not considered appropriately, the ISI can severely degrade the bit-error-ratio (BER) performance \cite{ jamali2016channel, fang2018symbol, jamali2018non}. Even though the use of enzymes \cite{noel2014improving} appropriately injected in the propagation environment can help mitigate the ISI, this approach cannot completely eliminate the ISI. Motivated by these considerations, we consider an MC system in the presence of ISI. Solutions to mitigate the ISI exist and include methods based on the design of the modulation \cite{mosayebi2018type}, the code \cite{shih2012channel}, and the detector \cite{kilinc2013receiver}. In this paper, we focus our attention on the design of detectors that use binary CSK modulation.

In MC systems, different types of detectors exist, including 1) no memory threshold-based schemes \cite{noel2014improving, damrath2016low}; 2) optimal constant-threshold-based schemes \cite{MDR_MolCOM2018}; 3) multi-memory-bit threshold-based schemes \cite{qian2019molecular}; 4) adaptive decision-feedback equalizers \cite{kilinc2013receiver}; and 5) multi-sub-slot threshold-based schemes \cite{mosayebi2014receivers}. Among these approaches, multi-memory-bit threshold-based schemes constitute a promising solution because they can achieve low BER performance at a low computational complexity. However, perfect CSI needs to be known at the receiver in order to calculate the optimal thresholds. In this regard, channel estimation (CE) \cite{jamali2016channel, noel2015joint} methods based on predefined preambles are often utilized in order to ensure the reliable detection of data. In realistic scenarios, however, some parameters of MC communication systems may not remain constant during the CE and data detection phases, thus leading to variations in the channel response and, hence, to CE inaccuracies. 
In some cases, e.g., for multiple absorbing receiver systems \cite{bao2019channel}, even though some theoretical channel models exist, which include MC systems with an eavesdropper \cite{GuoWEavesdropper}, MC systems with two receivers \cite{ XHuang2020, JWKwak2020} and MC systems with randomly distributed receivers \cite{SabuNV2020}, it may not be possible to model the channel analytically if realistic factors are considered.
To counteract these issues, some authors have used machine learning (ML) methods to design the receiver \cite{farsad2018sliding} and \cite{MDR_MolCOM2018}. However, ML-based receivers must be trained with long sequences of symbols. The performance of ML-based receivers, in addition, highly depends on the quality of the data used for training \cite{ zappone2019model}.

\subsection{Related Works}
Non-coherent detection methods for MC systems are discussed in \cite{jamali2018non}, where the authors obtain closed-form expressions of the optimal detection metric for a multi-symbol detector that uses an approximated probability mass function for the Poisson channel. In addition, the authors propose a blind single-symbol detector based on a constant threshold. However, the authors assume no ISI in the system model which may hold true when the symbol length is long enough and the symbol rate is low. 
In \cite{li2016local}, the authors consider the impact of ISI and exploit the local convexity of the received signals in order to detect the symbols from the difference of molecular concentration. To capitalize on the local convexity of the received signals, the transmitter needs, however, to release a sufficiently large number of molecules. In \cite{li2019csi}, a non-linear detector with an adaptive threshold is analyzed. The approach is based on the quick-rising and slow-decaying trends of the received signals after passing through a filter. In \cite{liu2020unsupervised}, the authors exploit the unsupervised fuzzy C-mean approach to detect symbols from the quick-rising and slow-decaying of the processed signals. 
In \cite{luo2019non}, the authors propose an approach based on the energy difference of the received signals. This method requires large numbers of molecules in order to exploit the energy difference. Non-coherent detection is investigated in Poisson channels for application to optical wireless communications in \cite{Gong2015}. However, current non-coherent detection schemes assume negligible ISI. 
In summary, the aforementioned approaches either do not consider ISI, or require large numbers of molecules and, in general, cannot achieve low BER performance.

\subsection{Contributions}
In this paper, motivated by the results in \cite{jamali2018non} and \cite{qian2019molecular}, we devise multi-memory-bit threshold-based schemes without prior CSI for the reliable transmission of data in the presence of ISI. 
The specific novelty and contributions made by this paper can be summarized as follows.
\begin{itemize}
\item We reformulate the multi-memory-bit threshold-based detectors via intermediate variables, i.e., the average number of received particles as a function of the memory bits. Unlike threshold-based methods that use the prior knowledge of the CSI, the considered intermediate variables can be obtained by applying clustering methods to the received signals. 
\item In order to reduce the clustering error, we construct multi-dimensional data from the received signals and designate the initial centroids of the clusters using the largest received signal. To the best of the authors’ knowledge, it is the first time in the communication literature that the main features are extracted by increasing the dimension for clustering data. Conventional methods, on the other hand, are based on decreasing the dimension for clustering data \cite{xu2005survey}. In addition, we can infer the transmitted symbols from the clustered data. We design and analyze two approaches, which are referred to as direct clustering-based inference and clustering-plus-threshold detection.
\item We devise an iterative method to further improve the BER performance and to avoid clustering failures in high data rate systems, i.e., in the presence of  non-negligible ISI. The essence of the proposed method is to iteratively apply clustering from one-dimensional data to high-dimensional data. In each iteration, we construct the initial centroids based on the estimated centroids in the previous iteration. By using the proposed iterative method, we show that the clustering errors can be reduced. 
\end{itemize}

\subsection{Paper Organization}
The remainder of this paper is organized as follows. In Section \ref{sec:system_model}, the system model is introduced. In Section \ref{sec:reformulation}, the multi-memory-bit threshold is reformulated via intermediate variables. In Section \ref{sec:non_coherent_approaches}, the proposed non-coherent detection method based on multi-dimensional clustering is introduced. In Section \ref{sec:non_coherent_approaches_iterative}, an iterative method for estimating the initial centroids for application to multi-dimensional clustering is introduced. 
In Section \ref{sec:results_analysis}, numerical results and the computational complexity of the proposed schemes are discussed. Finally, Section \ref{sec:conclusion} concludes this paper.

\section{System model}\label{sec:system_model}
\begin{figure}[!ht]
\centering\includegraphics[width=.5\textwidth]{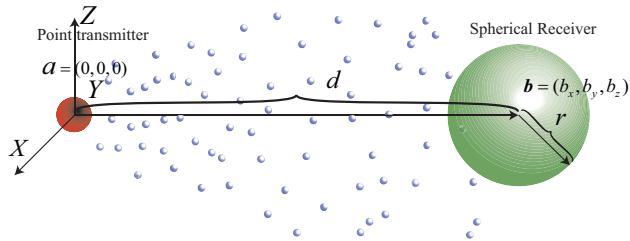}
\caption{Three-dimensional MC system consisting of a point transmitter and a spherical absorbing receiver  }\label{fig:threeD_MC_diagram}
\end{figure}

We consider a three-dimensional unbounded MC system without flow that consists of a point transmitter and a spherical absorbing receiver, as depicted in Fig. \ref{fig:threeD_MC_diagram}. We assume that each information particle diffuses randomly and independently through the medium. The particles are assumed not to degrade rapidly, thus resulting in ISI. We assume that the temperature is constant and the viscosity remains unchanged during the whole transmission. Thus, the diffusion coefficient $ D $ \cite{Hiyama2005} is assumed to be constant.

Assuming that the transmitter is located in $\bm{a} = (0,0,0)$ and the receiver is located in $\bm{b} = (b_x,b_y,b_z)$, the hitting rate is expressed as follows \cite{damrath2016low, yilmaz2014three}:
\begin{equation}\label{equ:hit_rate_3D}
f_{hit}(t)=\frac{r(d-r)}{d\sqrt{4\pi Dt^3}}e^{-\frac{(d-r)^2}{4Dt}}
\end{equation}
where $\|\bm{a}-\bm{b}\|=d$ is the distance between the center of the transmitter and the center of the receiver, and $r$ denotes the radius of the receiver. A binary CSK modulation scheme is assumed\footnote{ The generalization to multilevel CSK modulation is discussed in Section VI-C.}, and the bit transmitted in the $i$th time slot is denoted by $s_i$. During the $ i $th slot, the transmitter releases $ N_{TX} $ information particles if $ s_i=1 $, otherwise the transmitter does not release any particles. We assume that the $ N_{TX} $ information particles are released in a very short time such that we can ignore the release time effect on the received signal. The hitting probability of an absorbing receiver in the {\color{black}$(i-1)$}th time slot of duration $T$ is:
\begin{equation}
P_{i-1} = \int_{(i-1)T}^{iT}f_{hit}(t)dt = \frac{r}{d}\left\lbrace  \mathrm{erfc}(\frac{d-r}{\sqrt{4DiT}}) - \mathrm{erfc}(\frac{d-r}{\sqrt{4D(i-1)T}})\right\rbrace 
\label{equ:Prob_each_slot}
\end{equation}
where $ \mathrm{erf}(y)=\int_{0}^{y} \frac{2}{\sqrt{\pi }} e^{-x^2} dx $ and $ \mathrm{erfc}(y)=1-\mathrm{erf}(y) $.

Let $ C_j=N_{TX}P_j$ denote the average number of received particles in the $ j $th time-slot after the release of $ N_{TX} $ particles. {\color{black} For clarity, $C_0$ denotes the average number of received particles in the symbol of interest.} Thus, the number of received particles \cite{mosayebi2014receivers} in the $ i $th time-slot follows a Poisson distribution\footnote{ {\color{black} The number of received particles $r_i$ can be modelled accurately by using a Binomial distribution. For ease of analysis, $r_i$ is approximated with a Poisson distribution. This is a good approximation if $N_{TX}$ is large \cite{jamali2018non}. }  }:
\begin{equation}
   r_i\sim \mathrm{ Poisson }( I_i + s_iC_0)
   \label{equ:ri}
\end{equation}
where $ I_i = \bar{\lambda}_0 T + \sum\limits_{j=1}^L s_{i-j}C_j $ is the sum of the ISI and background noise, $ \bar{\lambda}_0 $ is the background noise power per unit time, and $L$ denotes the length (memory) of the Poisson channel.

We define the signal-noise-ratio (SNR)  as follows:
\begin{equation}
   \mathrm{SNR} =10 \log_{10} \frac{ C_0 }{2\bar{\lambda}_0T}
   \label{equ:SNR_defi}
\end{equation}
since the information bits are assumed to be equiprobable. Thus,
\begin{equation}
   N_{TX} = \frac{2{\color{black}\bar{\lambda}_0}T 10^{\frac{ \mathrm{SNR} }{10}}}{P_0}
   \label{equ:NTX_cal}
\end{equation}

Therefore, the probability of receiving $r_i$ information particles is:
\begin{equation}
   \mathrm{Pr}(r_i|I_i+s_iC_0)=\frac{e^{-(I_i + s_iC_0)}(I_i + s_iC_0)^{r_i}}{r_i!}
   \label{equ:r_probability}
\end{equation}

\section{Multi-memory-bit Threshold Reformulation}\label{sec:reformulation}
To detect the symbols, existing methods are based on calculating thresholds based on previously detected bits, which are referred to as memory bits. The main idea is to exploit previously detected bits, even erroneously estimated, in order to detect new transmitted bits based on the number of received particles. If the number of particles is below the threshold, a binary zero is estimated. Otherwise, a binary one is estimated. 
We first discuss the case of knowing the exact channel length $L$, and then we consider the case of limited channel information.
The analytical threshold based on the knowledge of $ L $ memory bits can be obtained from the following equation:
\begin{equation}
\mathrm{Pr}(r_i|s_i=0,s_{i-j},1\leq j\leq L)=\mathrm{Pr}(r_i|s_i=1,s_{i-j},1\leq j\leq L) \label{equ:Pr_s1_s0_long}
\end{equation}
where $ \mathrm{Pr}(r_i|s_i=0,s_{i-j},1\leq j\leq L) $ and $ \mathrm{Pr}(r_i|s_i=1,s_{i-j},1\leq j\leq L) $ denote the probabilities of receiving $ r_i $ information particles in the $ i $th time slot conditioned upon the previously transmitted symbols $ s_{i-j}$ for $1\leq j\leq L $ and the current symbol $ s_i=0 $ or $ s_i=1 $, respectively. These probabilities correspond to (\ref{equ:r_probability}).

By solving (\ref{equ:Pr_s1_s0_long}) based on (\ref{equ:r_probability}), the optimal threshold $ \tau|_{s_{i-j},1\leq j\leq L} $ conditioned on the previously transmitted symbols $ s_{i-j}$ for $1\leq j\leq L $ is as follows:
\begin{equation}
   \tau|_{s_{i-j},1\leq j\leq L} = \frac{C_0}{\ln(1+ \frac{C_0}{I_i})} = \frac{C_0}{\ln(1+ \frac{C_0}{ \bar{\lambda}_0T   + \sum_{j=1}^L {s}_{i-j}C_j})} \label{equ:threshold_optimal}
\end{equation}
Since the symbols $ s_{i-j} $ for $ 0\leq j\leq L $ are unknown, the previously estimated symbols $ \hat{s}_{i-j} $ for $ 0\leq j\leq L $ are used to compute the threshold in (\ref{equ:threshold_optimal}), i.e., $ \tau|_{\hat{s}_{i-j},1\leq j\leq L}= \frac{C_0}{\ln(1+ \frac{C_0}{ \bar{\lambda}_0T + \sum_{j=1}^L \hat{s}_{i-j}C_j })}$.
Based on the computed threshold, $ \hat{s}_{i} $ is demodulated as follows:
\begin{equation}
   \hat{s}_{i}=\left\{
   \begin{matrix}
     1 & r_i>\tau|_{\hat{s}_{i-j},1\leq j\leq L}\\
     0 & r_i\leq \tau|_{\hat{s}_{i-j},1\leq j\leq L}
\end{matrix} \right.
\end{equation}

Besides necessitating the bits detected in the previous time slots, the computation of the threshold in (\ref{equ:threshold_optimal}) relies upon the prior knowledge of the CSI, i.e., the coefficients $C_j$ are assumed to be known. To avoid using prior information about the channel, i.e., $ C_{j} $ for $ 0\leq j\leq L $, when calculating the threshold in (\ref{equ:threshold_optimal}), we resort to intermediate variables, i.e., the average number of received particles $ \bar{r}_i|_{s_{i-j},0\leq j\leq L} $ that is defined as follows:
\begin{equation}
   \bar{r}_i|_{s_{i-j},0\leq j\leq L} = I_i + s_iC_0= \bar{\lambda}_0T + \sum_{j=0}^L{s}_{i-j}C_j \label{equ:r_bar_L_exact}
\end{equation}

Equation (\ref{equ:r_bar_L_exact}) corresponds to the theoretical average number of particles, and, therefore, it depends on the variables $C_j$.
In the following, we first show how to formulate the thresholds in (8) by exploiting the intermediate variables in (10), instead of using $C_j$ explicitly. In this step the CSI is still required. Then, we elaborate how to estimate the intermediate variables without necessitating CSI.

\subsection{ 
Computing the Thresholds Using \texorpdfstring{$  \bar{r}_i|_{ s_{i-j},0\leq j\leq \mathcal{L}} $}{r_bar} } \label{sec:threshold_reformulation}
{\color{black} In this subsection, we show how the thresholds can be formulated only in terms of average number of particles $\bar{r}_i|_{s_{i-j},0\leq j\leq L}$.  }
In particular, in order to formulate the problem analytically and understand the rationale of the proposed approach, we consider, just for this section, that $\bar{r}_i|_{s_{i-j},0\leq j\leq L}$ is given by its analytical expression in (\ref{equ:r_bar_L_exact}).

The threshold in (\ref{equ:threshold_optimal}) can be rewritten, in an equivalent form, only in terms of $ \bar{r}_i|_{s_{i-j},0\leq j\leq L} $, as follows:
\begin{eqnarray}
   \tau|_{s_{i-j},1\leq j\leq L} &=& \frac{C_0}{\ln(1+ \frac{C_0}{ \bar{\lambda}_0T   + \sum_{j=1}^L {{s}}_{i-j}C_j})} 
   = \frac{ C_0+ \bar{\lambda}_0T + \sum_{j=1}^L {s}_{i-j}C_j - (  \bar{\lambda}_0T  +\sum_{j=1}^L {{s}}_{i-j}C_j)}{\ln( \frac{ \bar{\lambda}_0T  + C_0+ \sum_{j=1}^L {{s}}_{i-j}C_j}{ \bar{\lambda}_0T + \sum_{j=1}^L {{s}}_{i-j}C_j})} \nonumber \\
   &=& \frac{\bar{r}_i|_{s_i=1,s_{i-j},1\leq j\leq L}-\bar{r}_i|_{s_i=0,s_{i-j},1\leq j\leq L }}{\ln(\frac{\bar{r}_i|_{s_i=1,s_{i-j},1\leq j\leq L}}{\bar{r}_i|_{s_i=0,s_{i-j},1\leq j\leq L}} ) }
   \label{equ:threshold_derived}
\end{eqnarray}

In practice, the memory length used by the receiver may be limited and may be smaller than $L$. By assuming that $ \mathcal{L}\leq L $ memory bits are used, the threshold can be obtained by solving the equation $ \mathrm{Pr} (r_i|s_i=0,s_{i-j},1\leq j\leq \mathcal{L})=\mathrm{Pr} (r_i|s_i=1,s_{i-j},1\leq j\leq \mathcal{L}) $ where:
\begin{equation}
    \mathrm{Pr} (r_i|s_{i-j},0\leq j\leq \mathcal{L}) = \frac{1}{2^{L-\mathcal{L}}} \sum_{s_{i-j},\mathcal{L}+1\leq j\leq L} \mathrm{Pr} (r_i|s_{i-j},0\leq j\leq L) \label{equ:r_probability_cal_L_bits}
\end{equation}

Due to the analytical complexity of \eqref{equ:r_probability_cal_L_bits} it is, however, difficult to compute an analytical expression for the detection threshold by imposing $ \mathrm{Pr} (r_i|s_i=0,s_{i-j},1\leq j\leq \mathcal{L})=\mathrm{Pr} (r_i|s_i=1,s_{i-j},1\leq j\leq \mathcal{L})$. To overcome this analytical complexity, we use an approximation for $ \mathrm{Pr} (r_i|s_{i-j},0\leq j\leq \mathcal{L}) $. In particular, we approximate \eqref{equ:r_probability_cal_L_bits} with $ \mathrm{Pr}_{\mathrm{approx}} (r_i|s_{i-j},0\leq j\leq \mathcal{L}) $ as follows: 
\begin{equation}
   \mathrm{Pr}_{{\mathrm{approx}}} (r_i|s_{i-j},0\leq j\leq \mathcal{L})=\frac{e^{-\bar{r}_i|_{s_{i-j},0\leq j\leq \mathcal{L}}}(\bar{r}_i|_{s_{i-j},0\leq j\leq \mathcal{L}})^{r_i}}{r_i!}
   \label{equ:app_r_probability}
\end{equation}
which is obtained by averaging $ I_i = \bar{\lambda}_0 T + \sum\limits_{j=1}^L s_{i-j}C_j $ in (\ref{equ:r_probability}) with respect to $ s_{i-j}$ for $\mathcal{L}+1\leq j\leq L $, which yields:
\begin{equation}
   \bar{r}_i|_{s_{i-j},0\leq j\leq \mathcal{L}}= \sum_{j=0}^{\mathcal{L}}s_{i-j}C_j+\sum_{j=\mathcal{L}+1}^LC_j/2+\bar{\lambda}_0T \label{equ:r_bar_s_L}
\end{equation}

By using $ \mathrm{Pr}_{{\mathrm{approx}}} (r_i|s_{i-j},0\leq j\leq \mathcal{L}) $, the detection threshold can be computed by solving the equation $ \mathrm{Pr}_{\mathrm{approx}} (r_i|s_i=0,s_{i-j},1\leq j\leq \mathcal{L}) = \mathrm{Pr}_{\mathrm{approx}} (r_i|s_i=1,s_{i-j},1\leq j\leq \mathcal{L}) $, which yields:
\begin{equation}
   \tau|_{s_{i-j},1\leq j\leq \mathcal{L}} = \frac{\bar{r}_i|_{s_i=1,s_{i-j},1\leq j\leq \mathcal{L}}-\bar{r}_i|_{s_i=0,s_{i-j},1\leq j\leq \mathcal{L} }}{\ln(\frac{\bar{r}_i|_{s_i=1,s_{i-j},1\leq j\leq \mathcal{L}}}{\bar{r}_i|_{s_i=0,s_{i-j},1\leq j\leq \mathcal{L}}} ) }
   \label{equ:subopt_threshold_derived}
\end{equation}

As anticipated, the threshold in (\ref{equ:subopt_threshold_derived}) is a theoretical formulation for the time being. In practice, it needs to be obtained (estimated) only from the received data without any prior information. To this end, we replace $ \bar{r}_i|_{s_{i-j},0\leq j\leq \mathcal{L}} $ with its estimate $ \hat{\bar{r}}_i|_{ s_{i-j},0\leq j\leq \mathcal{L}} $, which is introduced and defined in the next sub-section. Therefore, the detection threshold is expressed as follows: 
\begin{equation}
   \hat{\bar{\tau}}|_{s_{i-j},1\leq j\leq \mathcal{L}} = \frac{\hat{\bar{r}}_i|_{s_i=1,s_{i-j},1\leq j\leq \mathcal{L}}- \hat{\bar{r}}_i|_{s_i=0,s_{i-j},1\leq j\leq \mathcal{L} }}{\ln(\frac{ \hat{\bar{r}}_i |_{s_i=1,s_{i-j},1\leq j\leq \mathcal{L}}}{ \hat{\bar{r}}_i|_{s_i=0,s_{i-j},1\leq j\leq \mathcal{L}}} ) }
   \label{equ:subopt_threshold_derived_practical}
\end{equation}

Since $ s_{i-j} $ for $ 0\leq j\leq \mathcal{L} $ are unknown, the estimates $ \hat{s}_{i-j} $ are used instead, which results in the following formulation for the threshold:
\begin{equation}
   \hat{\bar{\tau}}|_{\hat{s}_{i-j},1\leq j\leq \mathcal{L}} = \frac{\hat{\bar{r}}_i|_{\hat{s}_i=1,\hat{s}_{i-j},1\leq j\leq \mathcal{L}}-\hat{\bar{r}}_i|_{\hat{s}_i=0,\hat{s}_{i-j},1\leq j\leq \mathcal{L} }}{\ln(\frac{\hat{\bar{r}}_i|_{\hat{s}_i=1,\hat{s}_{i-j},1\leq j\leq \mathcal{L}}}{\hat{\bar{r}}_i|_{\hat{s}_i=0,\hat{s}_{i-j},1\leq j\leq \mathcal{L}}} ) }
   \label{equ:threshold_reformulated}
\end{equation}
Based on this reformulation, each symbol is demodulated as follows:
\begin{equation}
   \hat{s}_i = \left\{ \begin{array}{cl}
                0, & r_i\leq \hat{\bar{\tau}}|_{\hat{s}_{i-j},1\leq j\leq \mathcal{L}}   \\
                1, & r_i > \hat{\bar{\tau}}|_{\hat{s}_{i-j},1\leq j\leq \mathcal{L}}    \\
                \end{array}\right.
\label{equ:detection_with_threshold}             
\end{equation}

\subsection{Estimation of \texorpdfstring{$ \hat{\bar{r}}_i|_{ s_{i-j},0\leq j\leq \mathcal{L}} $}{r_bar_hat} from the Received Data}\label{sec:estimate_r_bar_hat}
As anticipated, our objective is to develop non-coherent detection schemes that do not need CSI to operate. For example, we cannot rely on the knowledge of $C_j$ for $ j=0,...,\mathcal{L} $ to compute the detection thresholds. In this sub-section, we show an example of how $ \hat{\bar{r}}_i|_{ s_{i-j},0\leq j\leq \mathcal{L}} $ can be obtained by only using the received data, which implies that the threshold is (\ref{equ:threshold_reformulated}) can be obtained without prior CSI information.

As a case study, we show how the intermediate variables $ \hat{\bar{r}}_i|_{ \hat{s}_{i-j},0\leq j\leq \mathcal{L}} $ are obtained for $\mathcal{L}=1$. { From (\ref{equ:r_probability_cal_L_bits}), the exact} probability mass function of receiving $ r_i $ particles given $ s_i$ and  $s_{i-1} $ is: 
\begin{equation}
   \mathrm{Pr}(r_i| s_i,s_{i-1} ) = \frac{1}{2^{L-1}}  \sum_{s_{i-j},2\leq j\leq L} \frac{e^{-(\bar{\lambda}_0 T + \sum\limits_{j=0}^L s_{i-j}C_j) } ( \bar{\lambda}_0 T + \sum\limits_{j=0}^L s_{i-j}C_j)^{r_i} }{r_i!}  \label{equ:pdf_r_s_i_s_i_1}
\end{equation} 
Theoretically, $ \bar{r}_i|_{s_i,s_{i-1} } $ is defined and can be computed as follows:
\begin{equation}
\bar{r}_i|_{s_i,s_{i-1}} = E[r_i|_{s_i,s_{i-1}}]  = \sum_{r_i=0}^{\infty} r_i \mathrm{Pr}(r_i| s_i,s_{i-1} ) \label{equ:r_bar_case_study_exact}
\end{equation}

Let us consider a sequence of received particles $ \left\lbrace r_1,...,r_n,...,r_K \right\rbrace  $ of length $ K $. The theoretical average number of received particles in (\ref{equ:r_bar_case_study_exact}) can be estimated empirically from the sequence of $ K $ observations. 
In practice (empirically), the probability $ \mathrm{Pr}(r_i| s_i,s_{i-1} ) $ can be interpreted as a ratio \cite[Section 19.2]{Van2000hAsymptoticBook}: 
\begin{equation}
\mathrm{Pr}(r_i| s_i,s_{i-1} ) = \lim_{ N_{s_i,s_{i-1}} \rightarrow \infty  } \frac{ N_{r_i,s_i,s_{i-1}} }{ N_{s_i,s_{i-1}}}
\end{equation}
where $ N_{r_i,s_i,s_{i-1}} $ denotes the number of elements within the sequence of $ K $ observations that are equal to $r_i$ and for which $[s_n,s_{n-1}]=[s_i,s_{i-1}]$, and $N_{s_i,s_{i-1}} = \sum_{r_i=0}^{\infty} N_{r_i,s_i,s_{i-1}} $ denotes the number of elements within the sequence of $ K $ observations for which $[s_n,s_{n-1}]=[s_i,s_{i-1}]$ (but the number of particles is not necessarily equal to $r_i$). Therefore, (\ref{equ:r_bar_case_study_exact}) can be rewritten as follows:  
\begin{eqnarray}
\bar{r}_i|_{s_i,s_{i-1}} = \lim_{ N_{s_i,s_{i-1}} \rightarrow \infty  }   \frac{ \sum_{r_i=0}^{\infty}r_i N_{r_i,s_i,s_{i-1}} }{ N_{s_i,s_{i-1}} } = \lim_{ K \rightarrow \infty  }   \frac{ \sum_{n=0}^{K}r_n \kappa_{n,s_i,s_{i-1}}  }{  \sum_{n=0}^{K} \kappa_{n,s_i,s_{i-1}}  } \label{equ:r_bar_case_study_new_form}
\end{eqnarray}
where $ \kappa_{n,s_i,s_{i-1}} = 1 $ if $ [s_n,s_{n-1}]=[s_i,s_{i-1}] $ and $ \kappa_{n,s_i,s_{i-1}} = 0 $ otherwise. 
The variable $ \kappa_{n,s_i,s_{i-1}} $ can be regarded as an indicator variable that provides information on whether the observation $ r_n $ belongs to a group of observations for which $ [s_n,s_{n-1}]=[s_i,s_{i-1}] $. 

In the literature \cite{jain1999data}, a group of observations is referred to as a cluster, a single observation of the cluster is referred to as a point of the cluster, and the condition that defines the cluster, i.e., the condition $ [s_n,s_{n-1}]=[s_i,s_{i-1}] $, is referred to as the label of the cluster.
In addition, the arithmetic mean of all the points in a subset $ X $ of $ \mathbb{R}^l $ is referred to as the centroid of the cluster \cite{bovik2010handbook} and is defined, in general terms, as follows:
\begin{equation}
\bm{\mu} = \frac{ \int \bm{x} g(\bm{x})d\bm{x} }{ \int g(\bm{x})d\bm{x} } \label{equ:centroid_definition_math}
\end{equation}
where $ \bm{x} $ is a vector in $ \mathbb{R}^l $, the integrals are computed over the whole space $ \mathbb{R}^l $, and $ g(\bm{x}) $ is a characteristic function that can be defined for different purposes. Based on (\ref{equ:r_bar_case_study_new_form}) and (\ref{equ:centroid_definition_math}), $ \bar{r}_i|_{s_i,s_{i-1}} $ can be viewed as the centroid of the cluster of observations whose label is $ [s_i,s_{i-1}] $. Thus $ \bar{r}_i|_{s_i,s_{i-1}} $ is obtained by computing the centroid of the cluster with label $ [s_i,s_{i-1}] $ that is obtained from the set of received signals (the observations) $ \left\lbrace r_1,...,r_n,...,r_K \right\rbrace  $.

Based on (\ref{equ:r_bar_case_study_new_form}), we evince that the thresholds of interest can be computed by calculating the centroids in (\ref{equ:centroid_definition_math}). However, (\ref{equ:r_bar_case_study_new_form}) necessitates the knowledge of the symbols $s_n$, which are usually unknown. 
Our objective is to estimate $ \bar{r}_i|_{s_i,s_{i-1}} $ from the received signals without knowing the symbols $ s_n $. Estimating the centroids in (\ref{equ:r_bar_case_study_new_form}) or (\ref{equ:centroid_definition_math}) without any prior information can be obtained by applying clustering methods, which partition the received signals into several groups, each one corresponding to a given label $ [s_i,s_{i-1}] $, and then obtaining the corresponding centroids.

In summary, the thresholds for data detection can be obtained by applying clustering methods directly to the received signals without any prior information on the CSI and without knowledge of the transmitted symbols. In particular, the thresholds can be retrieved from the centroids of the clusters. In the next section, we detail how to use clustering for non-coherent detection in MC systems, and, in particular, we introduce our approach based on multi-dimensional clustering.

\section{Clustering-based Non-coherent Detection}\label{sec:non_coherent_approaches}
With the aid of the reformulated thresholds in Section  \ref{sec:reformulation}, we show that data detection can be realized without prior CSI. In particular, we introduce non-coherent detection methods based on the K-means clustering algorithm. To this end, we first introduce some background information on clustering in general and the K-means algorithm in particular. To elucidate the operating principle of the proposed methods, we report some illustrations that are obtained by using the simulation setup in Table \ref{tab:parameter} (see Section \ref{sec:results_analysis}) in the presence of mild ISI ($T = 30 \Delta T$) and severe ISI ($T = 20 \Delta T$). It is worth mentioning that the algorithms proposed in this paper require the storage of symbols in order to be executed, similar to [16], [23], [27], [35]. At the time of writing, this may be difficult to be realized with biological circuits. The actual implementation of the proposed solutions by using biological circuits is an important issue that needs to be assessed in future research.

\subsection{K-Means Clustering Algorithm}
Assume that we have a data set $ \{\bm{x}_1,\bm{x}_2,...\bm{x}_N \} $ of $N$ observations and each element is a $ \mathbb{D} $-dimensional vector $ \bm{x}_n $. The objective is to partition the data set into $ N_c $ clusters whose centroids are $ \mathbb{D} $-dimensional vectors denoted by $ \bm{\mu}_k $. The centroid $ \bm{\mu}_k $ is the mean of its clustered points. In clustering, $ \kappa_{n,k} $ is a variable that represents if the distance between the $ n $th observation and the $ k $th centroid is smaller than the distance between that observation and the other centroids. 
The K-means algorithm is a clustering method that works iteratively in order to compute or estimate the centroids $\bm{\mu}_k$ and the indicator variables $\kappa_{n,k}$ from a set of observations. 
In particular, the K-means algorithm encompasses two steps that iteratively compute $\kappa_{n,k}$ and $\bm{\mu}_k$ at each step. The K-means algorithm needs an initial estimate of the centroids to operate. The initial centroids $ \bm{\mu}_k $ can be either randomly selected from the data set or other methods can be employed. The selection of the initial centroids is discussed in further text. The K-means clustering algorithm can be summarized as follows.

\textbf{Step I:}  Assign $ \bm{x}_n $ to the closest cluster (initial, if this is the first iteration) centroid: \\
\begin{equation}
   \kappa_{n,k} = \left\{ \begin{matrix}
     1, & if\quad k=\underset{j}{\arg\min} \, \Vert \bm{x}_n-\bm{\mu}_j \Vert^2  \\
     0, & otherwise
\end{matrix} \right.
\label{equ:assigning}
\end{equation}

\textbf{Step II:} Update the cluster centroid: \\
\begin{equation}
   \bm{\mu}_k = \frac{\sum_n \kappa_{n,k}\bm{x}_n}{\sum_n \kappa_{n,k}}
   \label{equ:centroid_update}
\end{equation}
where $ \sum_n \kappa_{n,k} $ is the number of points in the $k$th cluster.

The two steps are repeated until $ \kappa_{n,k} $ does not change. 
In the next sub-section, we show how to apply this algorithm for non-coherent detection in MC systems.

\subsection{Single-Dimensional Clustering: Challenges and Limitations}
In (\ref{equ:subopt_threshold_derived_practical}), the empirical average number of received particles $ \hat{\bar{r}}_i|_{s_{i-j},0\leq j\leq \mathcal{L}} $ need to be as close as possible to the theoretical average number of received particles $ \bar{r}_i|_{s_{i-j},0\leq j\leq \mathcal{L}} $ in order to obtain a low BER. In order to elucidate the complexity of the problem at hand, let us consider the example in (\ref{equ:pdf_r_s_i_s_i_1}) by assuming $\mathcal{L}=1$.

The objective is to estimate $ \hat{\bar{r}}_i|_{s_{i-j},0\leq j\leq \mathcal{L}} $ from the set of received observations, $ \left\lbrace r_1,...,r_n,...,r_K \right\rbrace  $, that are obtained from (\ref{equ:ri}) given the transmitted symbols $ \left\lbrace s_1,...,s_n,...,s_K \right\rbrace  $. Since $\mathcal{L}=1$, we are interested in identifying four clusters in the received signals, which correspond to the labels $[s_i,s_{{\color{black}i-1}}]=[0,0]$, $[s_i,s_{{\color{black}i-1}}]=[0,1]$, $[s_i,s_{{\color{black}i-1}}]=[1,0]$, and $[s_i,s_{{\color{black}i-1}}]=[1,1]$. 
From (\ref{equ:r_bar_case_study_exact}) or equivalently (\ref{equ:r_bar_s_L}), we obtain:
\begin{equation}
\bar{r}_i|_{s_i,s_{i-1}}=  s_iC_0+s_{i-1}C_1+\sum_{j=2}^LC_j/2 + \bar{\lambda}_0T \label{equ:r_bar_s_i_s_i_1_case_study_cal}
\end{equation}

\begin{figure}[!t]
\centering\includegraphics[width=0.7\textwidth]{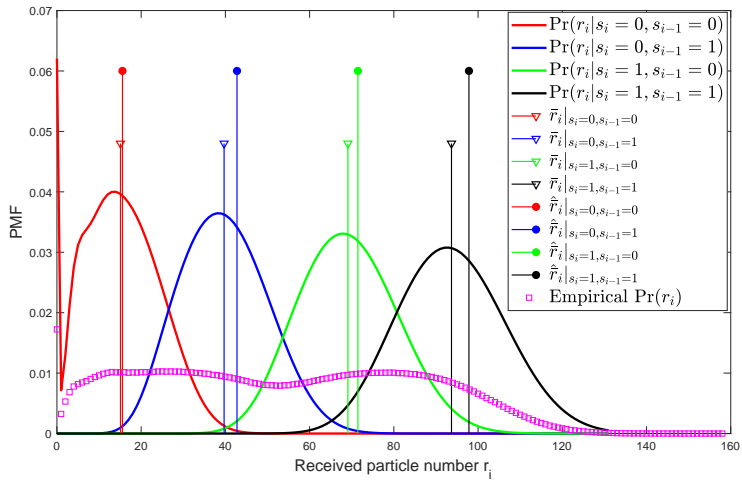}
\caption{The positions of $ \bar{r}_i|_{s_i,s_{i-1}} $ (calculated via (\ref{equ:r_bar_s_i_s_i_1_case_study_cal})) and $ \hat{\bar{r}}_i|_{s_i,s_{i-1}} $ corresponding to the centroids calculated from (\ref{equ:assigning}) and (\ref{equ:centroid_update}) and assuming that the labels $ [s_i,s_{i-1}] $ are known) for $T = 30 \Delta T$ }
\label{fig:distribution_overlap_more}
\end{figure}

The average number of received particles in (\ref{equ:r_bar_s_i_s_i_1_case_study_cal}) is the theoretical one that can be obtained from analysis. In order to understand the difficulty of obtaining (\ref{equ:r_bar_s_i_s_i_1_case_study_cal}) from the empirical data, we illustrate an example in Fig. \ref{fig:distribution_overlap_more}. Based on a large set of empirical data ($K=2^{16}$ samples), we calculate the empirical probability mass function (PMF) of the number of particles, by assuming that the labels $[s_i, s_{i-1}]$ are known. How the labels can be estimated from the data is discussed in further text. From the empirical data, we apply the K-means clustering algorithm in (\ref{equ:assigning}) and (\ref{equ:centroid_update}) in order to estimate the centroids. The estimated average number of received particles $\hat{\bar{r}}_i|_{s_i,s_{i-1}}$ is set equal to the estimated centroids. From Fig. \ref{fig:distribution_overlap_more}, we observe that there is a non-negligible difference between the theoretical values obtained from (\ref{equ:r_bar_s_i_s_i_1_case_study_cal}) and the empirical values estimated by using the K-means clustering algorithm, even if the labels are assumed to be known, i.e., the association between the centroids and the labels is error-free. The differences between $ \hat{\bar{r}}_i|_{s_i,s_{i-1} } $ and $ \bar{r}_i|_{s_i,s_{i-1} } $ mainly originate from the overlapping areas of the conditional probabilities $ \mathrm{Pr}(r_i| s_i,s_{i-1} ) $.

The simple example illustrated in Fig. \ref{fig:distribution_overlap_more} highlights the challenges of applying clustering for estimating the average number of received particles in MC systems and, therefore, the detection thresholds. In the following sub-section, we introduce our proposed approach for improving the estimation accuracy based on multi-dimensional clustering methods.

\subsection{Multi-Dimensional Clustering}
The proposed solution to increase the accuracy of estimating the average number of received particles from the empirical data relies on multi-dimensional clustering. The main idea is not to apply the K-means clustering algorithm in (\ref{equ:assigning}) and (\ref{equ:centroid_update}) to individual observations $r_n$ but to vectors of observations:
\begin{equation}
\bm{r}_n = [ r_n, r_{n-1}, ..., r_{n-\mathcal{L}} ] \label{equ:multi_dimension_data_construction}
\end{equation}
where $\mathcal{L}$ is the number of memory bits that we can use for detection and for multi-dimensional clustering. In order to illustrate the advantages of the proposed approach, let us consider the same example as in Fig. \ref{fig:distribution_overlap_more}, but by assuming that the K-means clustering algorithm in (\ref{equ:assigning}) and (\ref{equ:centroid_update}) is applied to the vector $[r_n, r_{n-1}]$, rather than to $r_n$ only.

\begin{figure}[!t]
\centering 
\subfigure[Empirical distributions of \texorpdfstring{$ [r_n,r_{n-1} ]|_{s_i,s_{i-1}} $}{original} and theoretical centroids obtained from (\ref{equ:2D_prac_centers}). In this case, the labels are assumed to be known a priori]{\label{fig:2_D_r_distribution_compa_long_a} \includegraphics[width=0.45\textwidth]{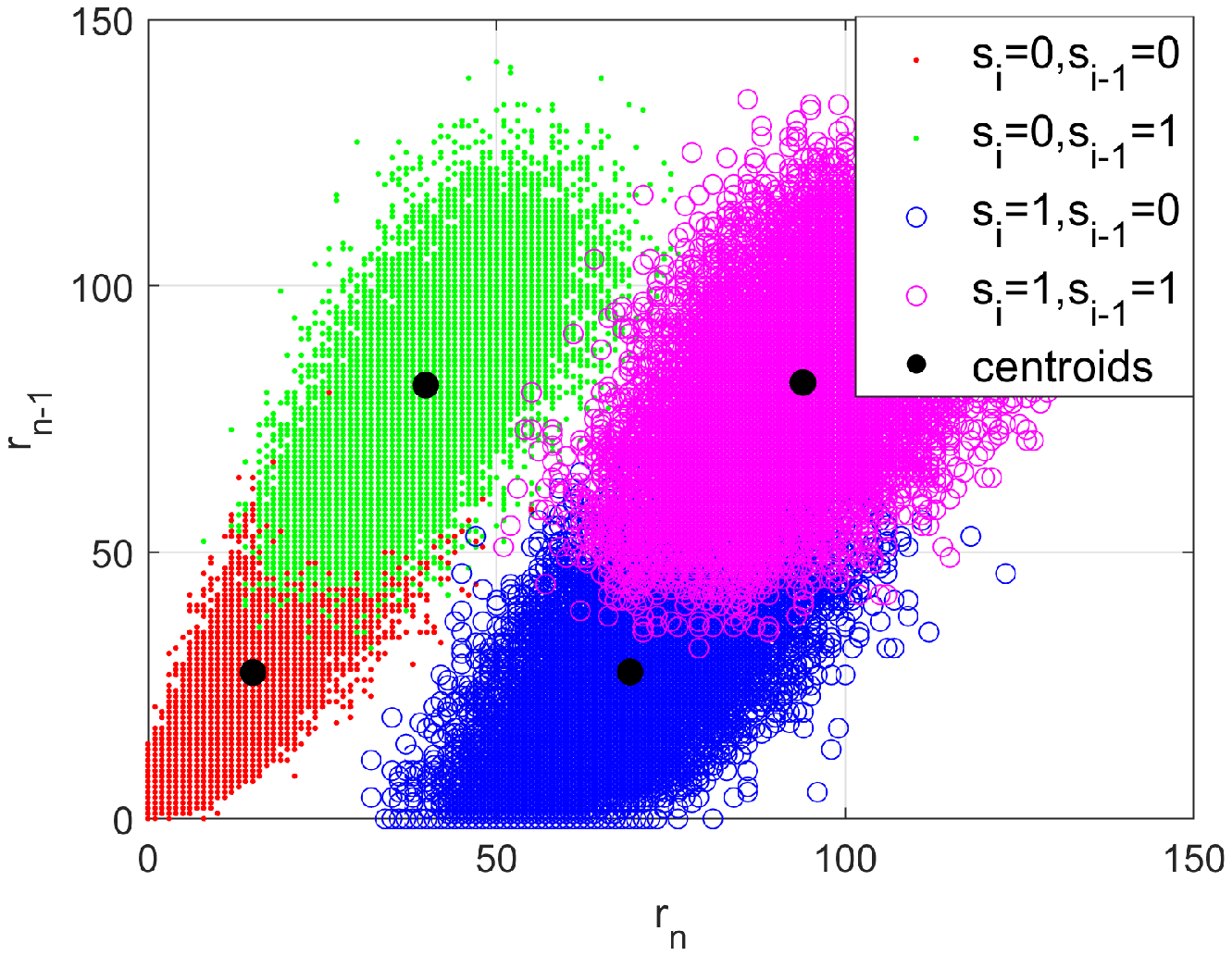}}
\subfigure[Theoretical centroids obtained by using (\ref{equ:2D_prac_centers}) (yellow circles) vs. estimated centroids obtained by using the K-means algorithms in (\ref{equ:assigning}) and (\ref{equ:centroid_update}) (blue circles). The labels are not known a priori but are obtained from the estimated centroids (the updated initial centroids with assigned labels)]{\label{fig:2_D_r_distribution_compa_long_b} \includegraphics[width=0.45\textwidth]{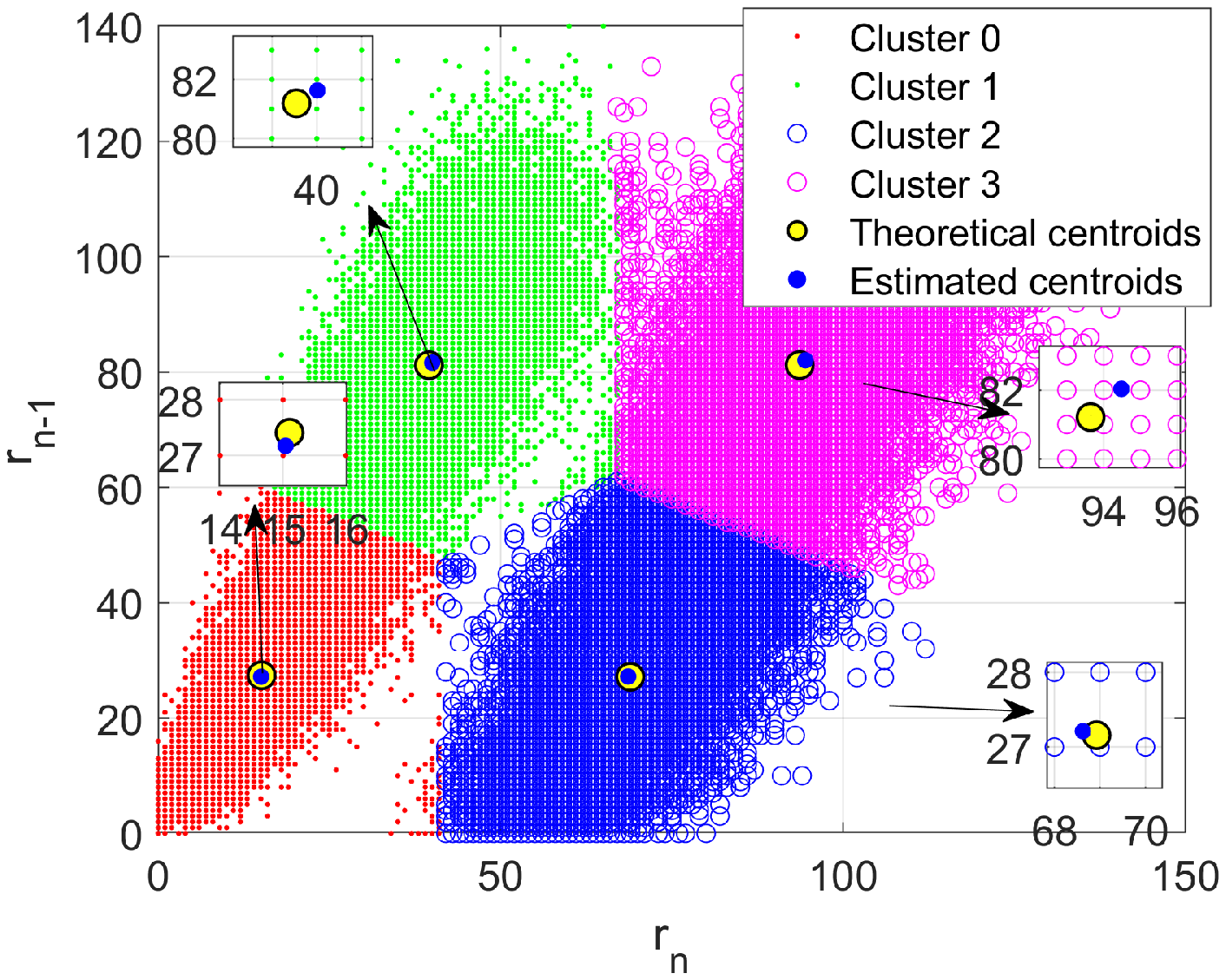}}
\caption{Theoretical and estimated centroids obtained from clustering the data vector $ [r_n,r_{n-1}] $ ($T = 30 \Delta T$)}
\label{fig:2_D_r_distribution_compa_long}
\end{figure}

The results are illustrated in Fig. \ref{fig:2_D_r_distribution_compa_long}, where the following notation is used. The cluster of points whose label is, e.g., $ [s_i=0,s_{i-1}=1] $ is denoted by $ [r_n,r_{n-1} ]|_{s_i=0,s_{i-1}=1} $. A similar notation is used for the other clusters. 
By definition, $ \bar{r}_i|_{s_i,s_{i-1}}=s_iC_0+s_{i-1}C_1+\sum_{j=2}^LC_j/2 + \bar{\lambda}_0T $ and $ \bar{r}_{i-1}|_{s_{i-1}}=s_{i-1}C_0+\sum_{j=1}^LC_j/2  + \bar{\lambda}_0T$ given the label $ [s_i,s_{i-1}] $. Therefore, the corresponding theoretical centroids are computed from (\ref{equ:r_bar_s_L}) as follows: 
\begin{equation}
\begin{split}
[\bar{r}_i,\bar{r}_{i-1} ]|_{s_i=0,s_{i-1}=0} & = [\sum_{j=2}^LC_j/2 , \sum_{j=1}^LC_j/2 ] + [\bar{\lambda}_0T,\bar{\lambda}_0T]\\
[\bar{r}_i,\bar{r}_{i-1} ]|_{s_i=0,s_{i-1}=1} & = [\sum_{j=2}^LC_j/2, \sum_{j=1}^LC_j/2 ]  +[C_1+\bar{\lambda}_0T,C_0+\bar{\lambda}_0T]\\
[\bar{r}_i,\bar{r}_{i-1} ]|_{s_i=1,s_{i-1}=0} & = [\sum_{j=2}^LC_j/2 , \sum_{j=1}^LC_j/2 ] +[C_0+\bar{\lambda}_0T,\bar{\lambda}_0T]\\
[\bar{r}_i,\bar{r}_{i-1} ]|_{s_i=1,s_{i-1}=1} & = [\sum_{j=2}^LC_j/2 , \sum_{j=1}^LC_j/2 ] +[C_0+C_1+\bar{\lambda}_0T,C_0+\bar{\lambda}_0T] 
\end{split} \label{equ:2D_prac_centers}
\end{equation}

Given a sequence of received signals (observations), $ \left\lbrace r_1, ..., r_n,..., r_K \right\rbrace  $, the vectors of observations are $ \left\lbrace [r_2,r_1], ..., [r_n,r_{n-1}],..., [r_K,r_{K-1}] \right\rbrace  $. The empirical distributions of each cluster of points and the theoretical centroids (computed from (\ref{equ:2D_prac_centers})) are depicted in Fig. \ref{fig:2_D_r_distribution_compa_long_a}. We apply the K-means clustering algorithm in (\ref{equ:assigning}) and (\ref{equ:centroid_update}) to the empirical vectors using some initial centroids  $ \bm{\mu}_k $.  The setup of the initial centroids is described in the next sub-section. In general, the exact labels $[s_n, s_{n-1}]$ are unknown, which implies that the association between an estimated centroid and the corresponding label $[s_i, s_{i-1}]$ is not known a priori. We solve this issue by appropriately selecting the initial centroids, and by deciding the association between centroids and labels at the beginning of the K-means algorithm. Based on our approach, the estimated final centroids inherit the labels associated to the initial centroids. If $[s_i, s_{i-1}]$ is the label of the initial centroid $\bm{\mu}_k$, then $[s_i, s_{i-1}]$ will be the label of the estimated (final) centroid $\hat{\bm{\mu}}_k$. Therefore, the choice of the initial centroids is important to ensure a correct labelling. By comparing Fig. \ref{fig:2_D_r_distribution_compa_long_b} and Fig. \ref{fig:distribution_overlap_more}, we observe that $ \hat{\bar{r}}_i|_{s_i,s_{i-1}} $ is closer to $ \bar{r}_i|_{s_i,s_{i-1}} $ in Fig. \ref{fig:2_D_r_distribution_compa_long_b}, which results in better performance. Therefore, multi-dimensional clustering is shown to be more accurate than one-dimensional clustering.

\subsection{Setup of the Initial Centroids}

\begin{figure}[!t]
\centering 
\subfigure[Initial centroids using \texorpdfstring{$ \max(\bm{r}) $}{original} ]{\label{fig:kmeans_direct_updated_centroids} \includegraphics[width=0.42\textwidth]{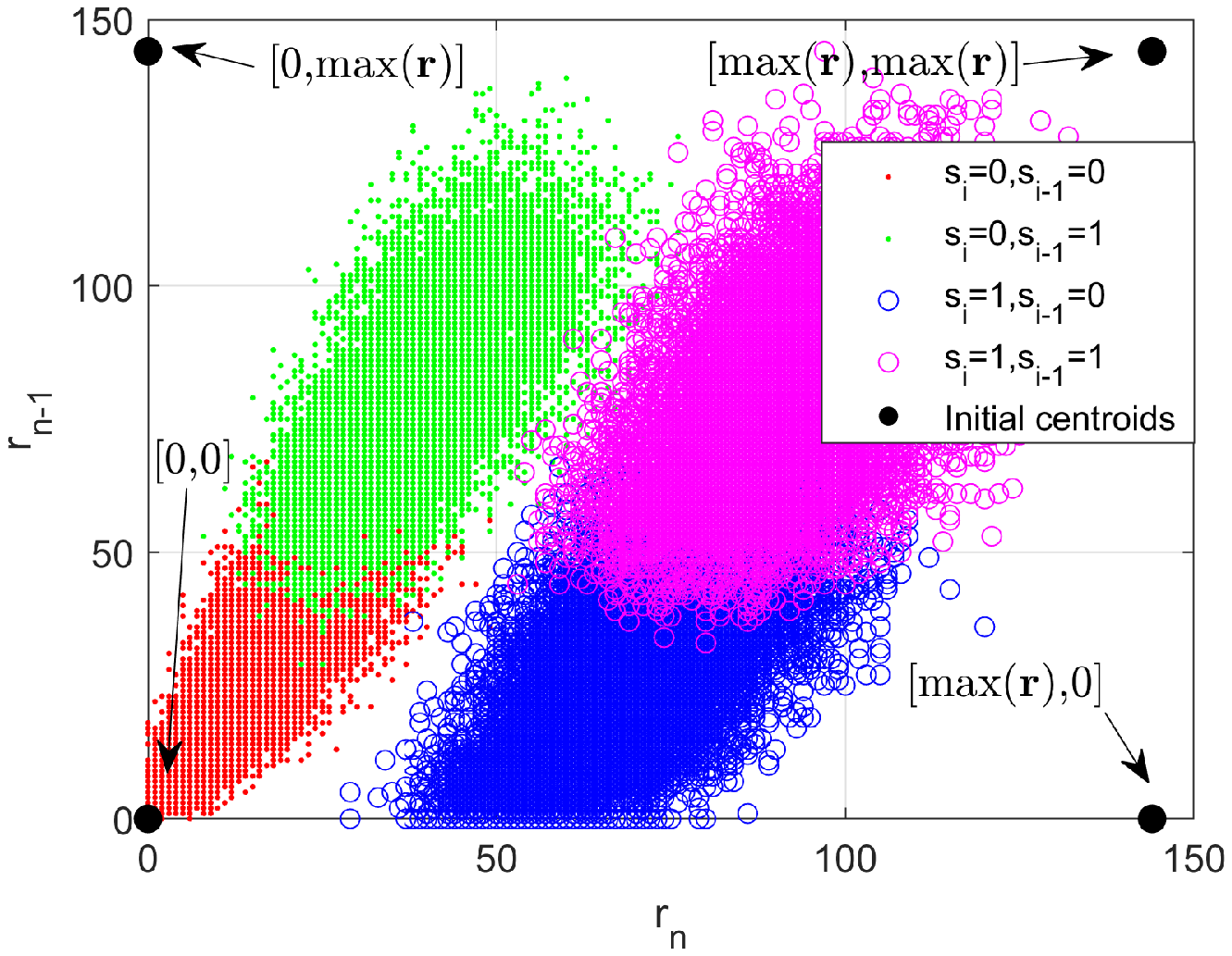}}
\subfigure[Initial centroids using \texorpdfstring{$ \hat{\bar{r}}_i|_{s_i} $}{original}]{\label{fig:kmeans_iter_updated_centroids} \includegraphics[width=0.45\textwidth]{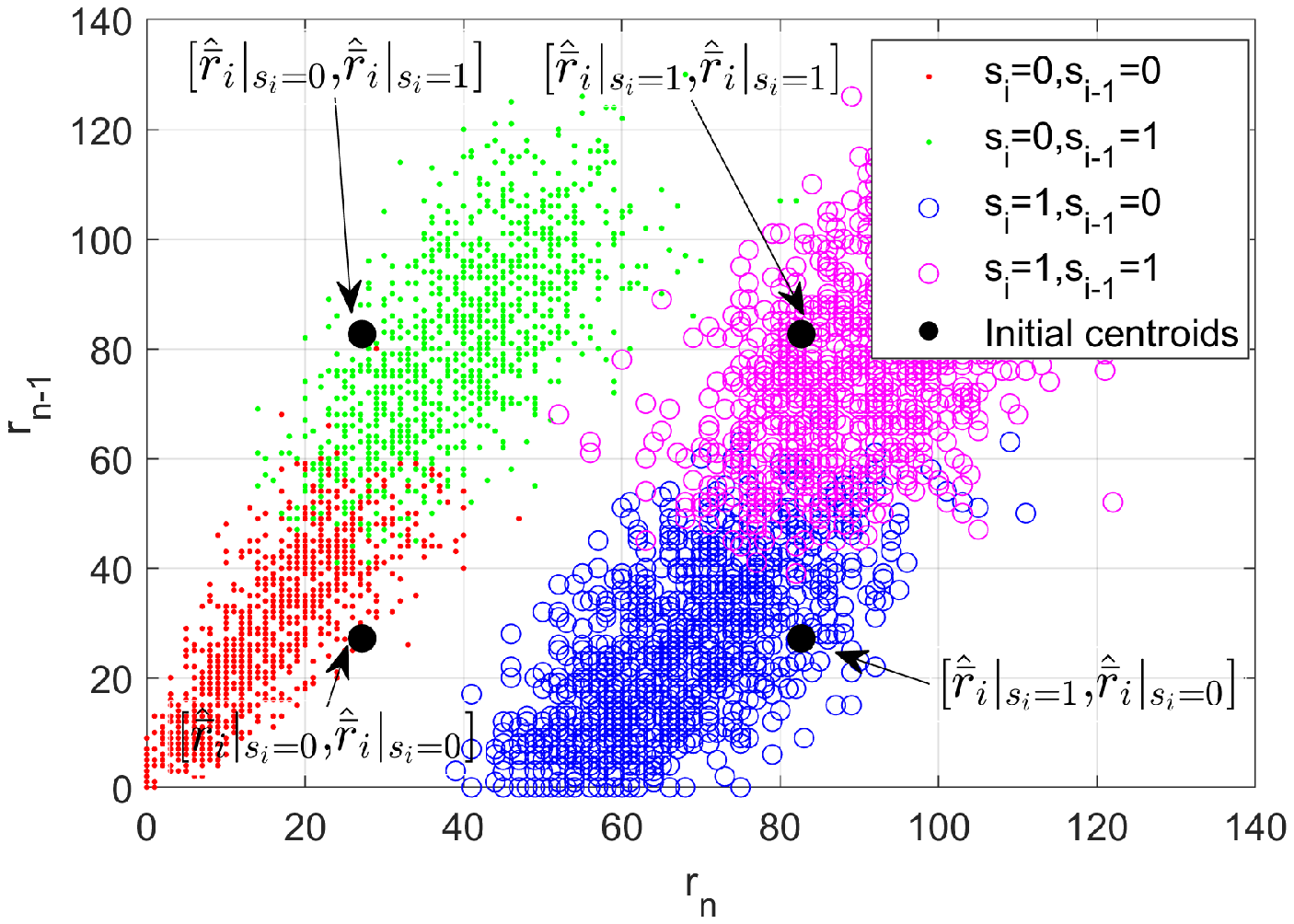}}
\caption{The customized initial centroids and the corresponding clusters ($T = 30 \Delta T$)}
\label{fig:kmeans_centroids}
\end{figure}

As mentioned, an important issue is to assign the correct labels to the initial centroids and to ensure that the association between the estimated centroids and the correct labels does not change when applying the K-means clustering algorithm. In order to motivate the proposed approach for the setup of the initial centroids and the corresponding labels, let us consider Fig. \ref{fig:kmeans_direct_updated_centroids}. In the figure, the four black dots correspond to the points $[0,0]$, $[0,\max(\bm{r})]$, $[\max(\bm{r}),0]$, and $[\max(\bm{r}), \max(\bm{r})]$, where $\max(\bm{r})$ returns that largest value of the observations $\{r_1, ....r_n....r_K\}$. {\color{black} In our proposed approach, the initial centroids of the clusters are chosen to correspond to the four black dots in Fig. \ref{fig:kmeans_direct_updated_centroids}.  The rationale for this choice lies in the fact that the maximum received signal can be readily computed from the empirical data. In addition, more importantly, due to the geometry of the problem at hand, our proposed choice leads to a low probability that the label of the centroids will change after the application of the clustering algorithm}. By direct inspection of the figure, in fact, the four black dots provide us with the correct label of the centroids, i.e., $[0,0]$ corresponds to the label $[s_i,s_{i-1}]=[0,0]$, $[0,\max(\bm{r})]$ corresponds to the label $[s_i,s_{i-1}]=[0,1]$, etc. If $K$ is sufficiently large, the proposed association between centroids and labels is expected not to change, with high probability, during the application of the K-means algorithm.

If the number of memory-bits $\mathcal{L}$ is greater than one, the initial centroids are assigned in a similar fashion. If $\mathcal{L}=2$, for example, the initial centroids are $[0,0,0]$ whose label is $[s_i, s_{i-1}, s_{i-2}]=[0,0,0]$; $[0,0,\max(\bm{r})]$ whose label is $[s_i, s_{i-1}, s_{i-2}]=[0,0,1]$; $[0,\max(\bm{r}),\max(\bm{r})]$ whole label is $[s_i, s_{i-1}, s_{i-2}]=[0,1,1]$; etc.

Based on these proposed initial centroids and the corresponding labels, the K-means clustering algorithm is applied according to (\ref{equ:assigning}) and (\ref{equ:centroid_update}), which returns the final estimated centroids ($\hat{\bm{\mu}}_k$) and the indicator variables $\kappa_{n,k}$. In particular, $\kappa_{n,k}$ allows us to implicitly perform data detection at the end of the clustering process, since it informs us, by definition, if a vector of points belongs or not to a given cluster. Based on this remark, we propose two algorithms for data detection that are referred to as (1) direct clustering-based inference and (2) clustering-plus-threshold detection.

\subsection{Direct Clustering-Based Inference and Clustering-plus-Threshold Detection}

The proposed \textit{direct clustering-based inference} algorithm is based only on clustering methods and does not exploit the thresholds in (\ref{equ:threshold_reformulated}) for data detection. The algorithm is given in Algorithm \ref{alg:clustering_based_inference} and works as follows (assuming, e.g., $\mathcal{L}=1$).
If the observation vector $[r_n, r_{n-1}]$ belongs to the cluster with label $[s_i, s_{i-1}]$, i.e., the corresponding indicators variables are $\kappa_{n,k}=1$, then the estimated bits are those of the corresponding label. 

The proposed \textit{clustering-plus-threshold detection} algorithm, on the other hand, combines together clustering methods and the estimated thresholds in (\ref{equ:threshold_reformulated}) for data detection. The algorithm is given in Algorithm \ref{alg:clustering_plus_threshold} and it has one main difference compared with Algorithm \ref{alg:clustering_based_inference}: After finalizing the clustering process based on the K-means algorithm, the bits are not detected by using the indicator variables $\kappa_{n,k}$, but the thresholds in (\ref{equ:threshold_reformulated}) are computed from the estimated centroids, which are then used for data detection by using (\ref{equ:detection_with_threshold}).

\begin{figure*}[!t]
\centering
\begin{minipage}[t]{.75\textwidth}
\begin{algorithm}[H]
 \caption{ { Direct clustering-based inference } }
 \begin{algorithmic}[1]
 \renewcommand{\algorithmicrequire}{\textbf{First Step:}}
 \renewcommand{\algorithmicensure}{\textbf{Second Step:}}
 \REQUIRE Clustering
  \STATE Set $ \mathcal{L} $
  \STATE Construct the data $ \bm{r}_n  $ via  (\ref{equ:multi_dimension_data_construction}) 
  \STATE Construct the initial centroids $ \bm{\mu}_k $ 
  \STATE Cluster the $ \bm{r}_n  $ using the K-means algorithm with the initial centroids $ \bm{\mu}_k $ 
 \ENSURE  Inference
  \STATE Infer $ \hat{s}_n $ from the indicator variables $ \kappa_{n,k} $
 \end{algorithmic} 
 \label{alg:clustering_based_inference}
\end{algorithm}
\end{minipage}
\begin{minipage}[t]{.75\textwidth}
\begin{algorithm}[H]
 \caption{{Clustering-plus-threshold detection}}
 \begin{algorithmic}[1]
 \renewcommand{\algorithmicrequire}{\textbf{First Step:}}
 \renewcommand{\algorithmicensure}{\textbf{Second Step:}}
 \REQUIRE Clustering
  \STATE Set $ \mathcal{L} $
  \STATE Construct the data $ \bm{r}_n  $ via  (\ref{equ:multi_dimension_data_construction}) 
  \STATE Construct the initial centroids $ \bm{\mu}_k $ 
  \STATE Cluster the $ \bm{r}_n  $ using the K-means algorithm with the initial centroids $ \bm{\mu}_k $  
 \ENSURE  Detection
  \STATE Obtain $ \hat{\bar{r}}_i|_{\hat{s}_{i-j},0\leq j\leq \mathcal{L}} $ from the estimated centroids $ \hat{\bm{\mu}}_k $ 
  \STATE Compute the detection thresholds using (\ref{equ:threshold_reformulated})
  \STATE Detect the symbols using (\ref{equ:detection_with_threshold}) 
 \end{algorithmic} 
 \label{alg:clustering_plus_threshold}
\end{algorithm}
\end{minipage}
\end{figure*}

\begin{figure}[!t]
\centering
\subfigure[Transmitted bits and exact labels]{\label{fig:simple_cluster_2D_w_tau_real} \includegraphics[width=0.45\textwidth]{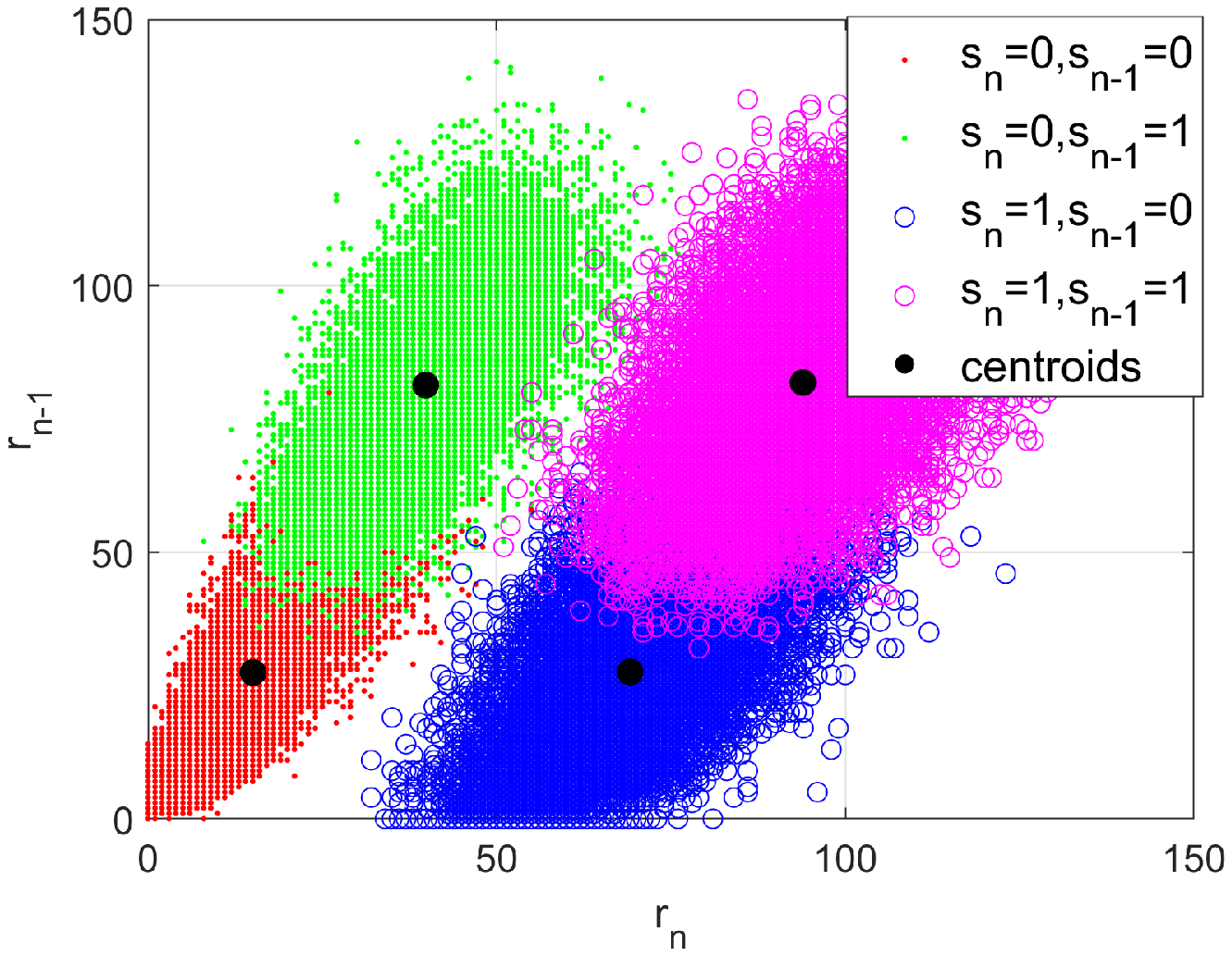}}
\subfigure[Direct clustering-based inference]{\label{fig:simple_cluster_2D_w_tau_a} \includegraphics[width=0.45\textwidth]{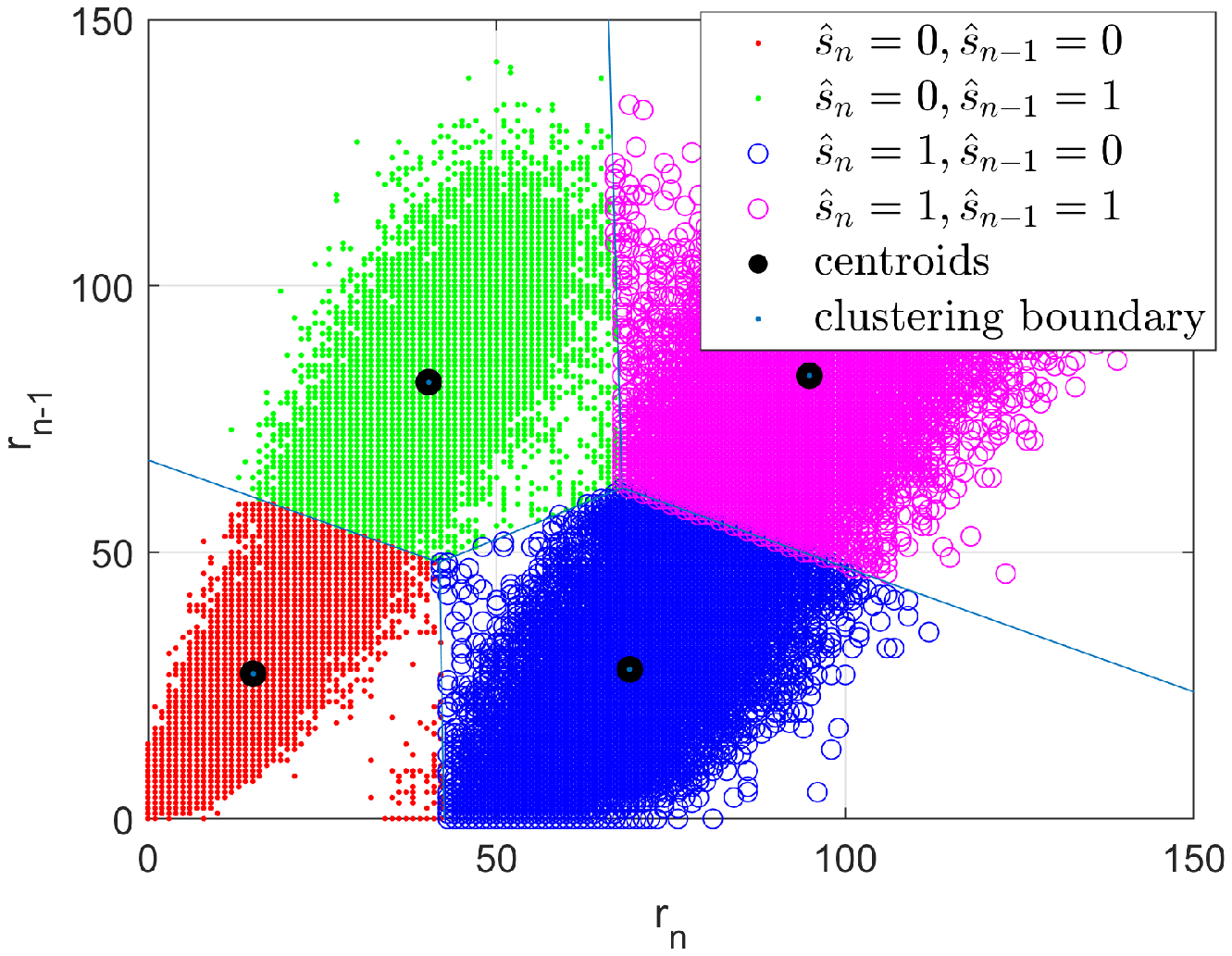}}
\subfigure[Clustering-plus-threshold detection ]{\label{fig:simple_cluster_2D_w_tau_b} \includegraphics[width=0.45\textwidth]{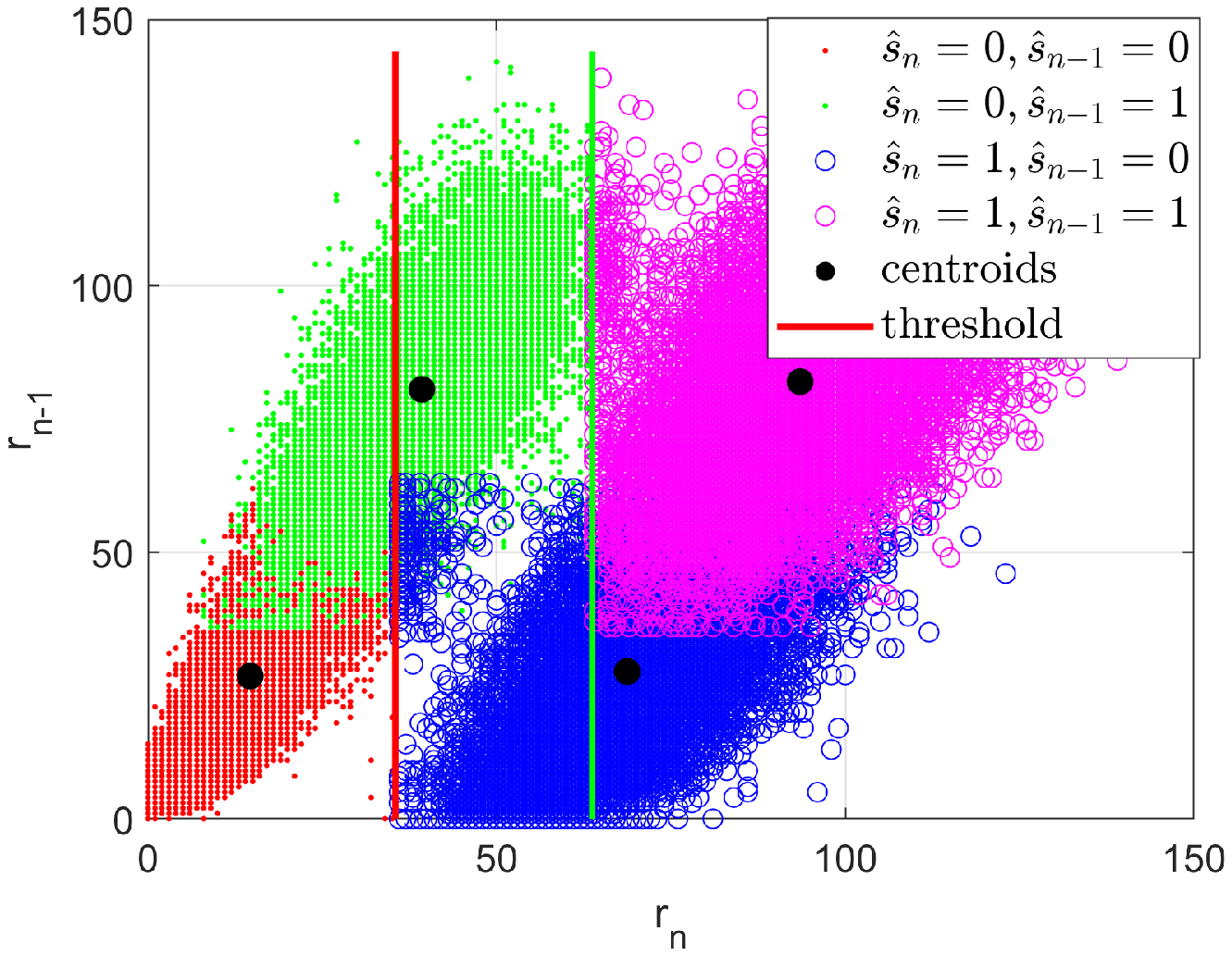}}
\caption{Distribution of $ [r_n,r_{n-1}]|_{\hat{s}_n,\hat{s}_{n-1}} $, estimated centroids and clustering boundaries/thresholds  ($ \mathcal{L}=1 $ and $T=30\ \Delta T$). }
\label{fig:simple_cluster_2D_w_tau}
\end{figure}

\begin{figure}[!t]
\centering
\subfigure[Direct clustering-based inference]{\label{fig:simple_cluster_5D_w_tau_a} \includegraphics[width=0.45\textwidth]{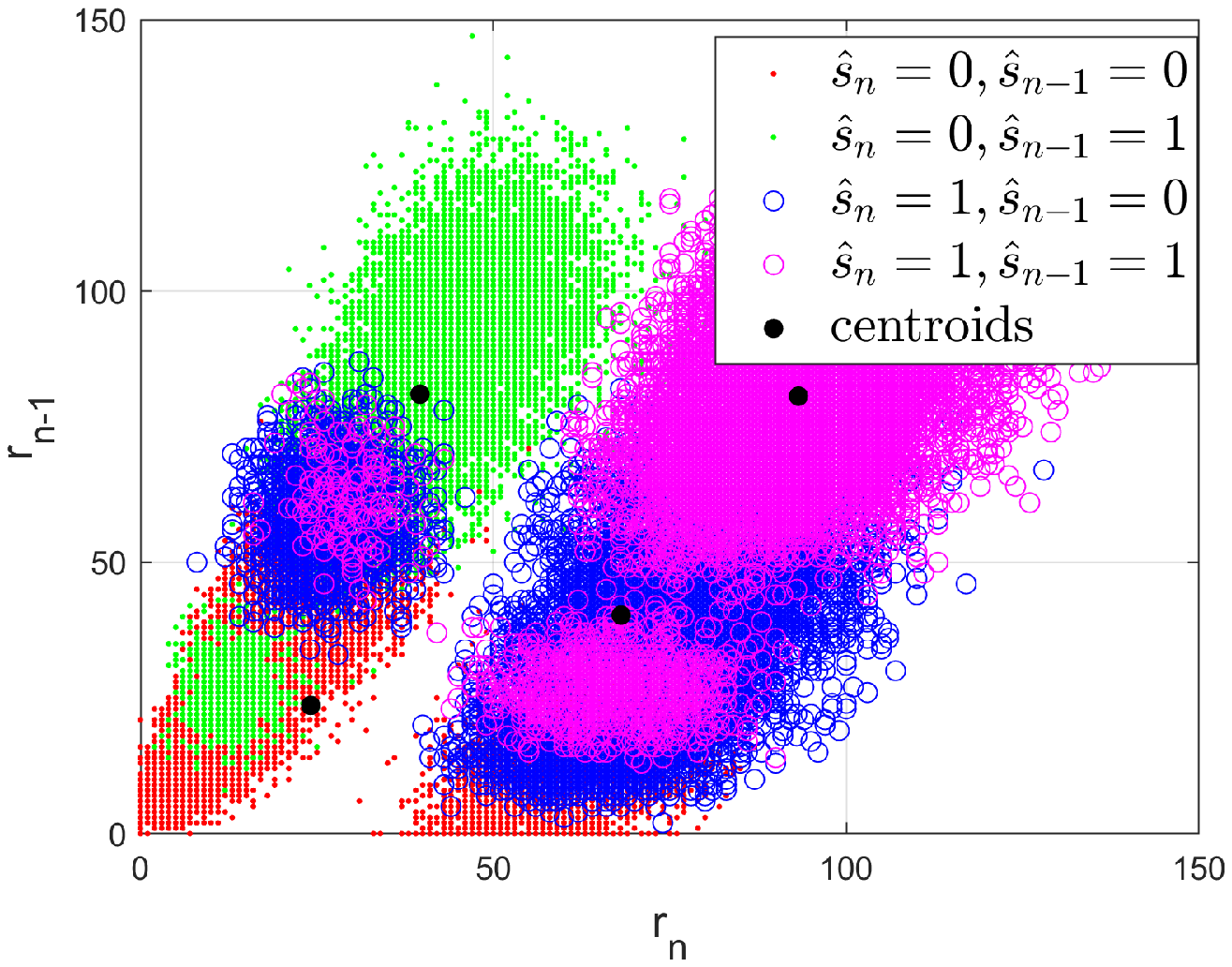}}
\subfigure[Clustering-plus-threshold detection]{\label{fig:simple_cluster_5D_w_tau_b} \includegraphics[width=0.45\textwidth]{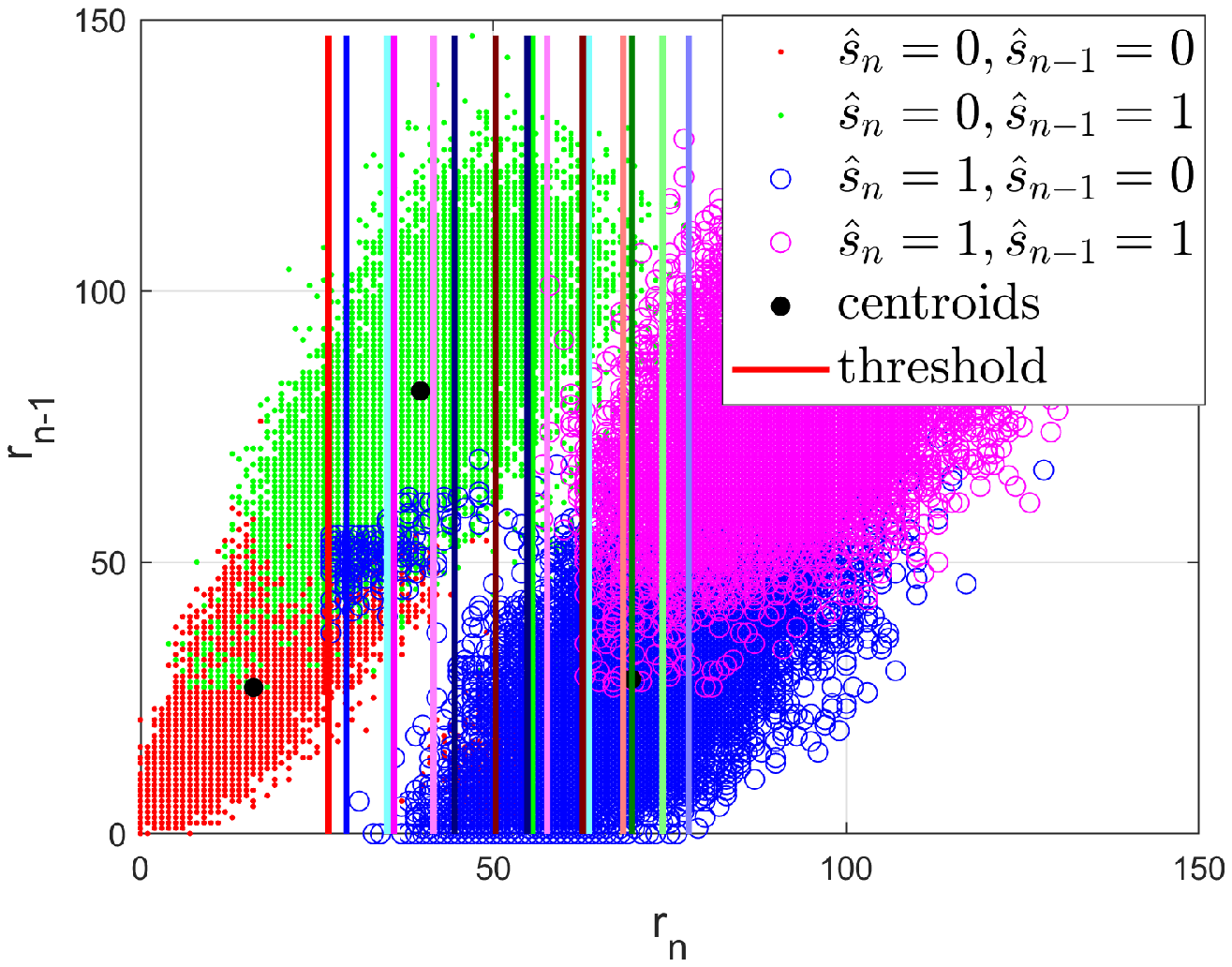}}
\caption{Distribution of $ [r_n,r_{n-1}]|_{\hat{s}_n,\hat{s}_{n-1}} $ and estimated centroids  ($ \mathcal{L}=4 $ and $T=30\ \Delta T$). } 
\label{fig:simple_cluster_5D_w_tau}
\end{figure}

\begin{figure}[!t]
\centering
\subfigure[Transmitted bits and exact labels]{\label{fig:simple_cluster_2D_w_tau_real_short} \includegraphics[width=0.45\textwidth]{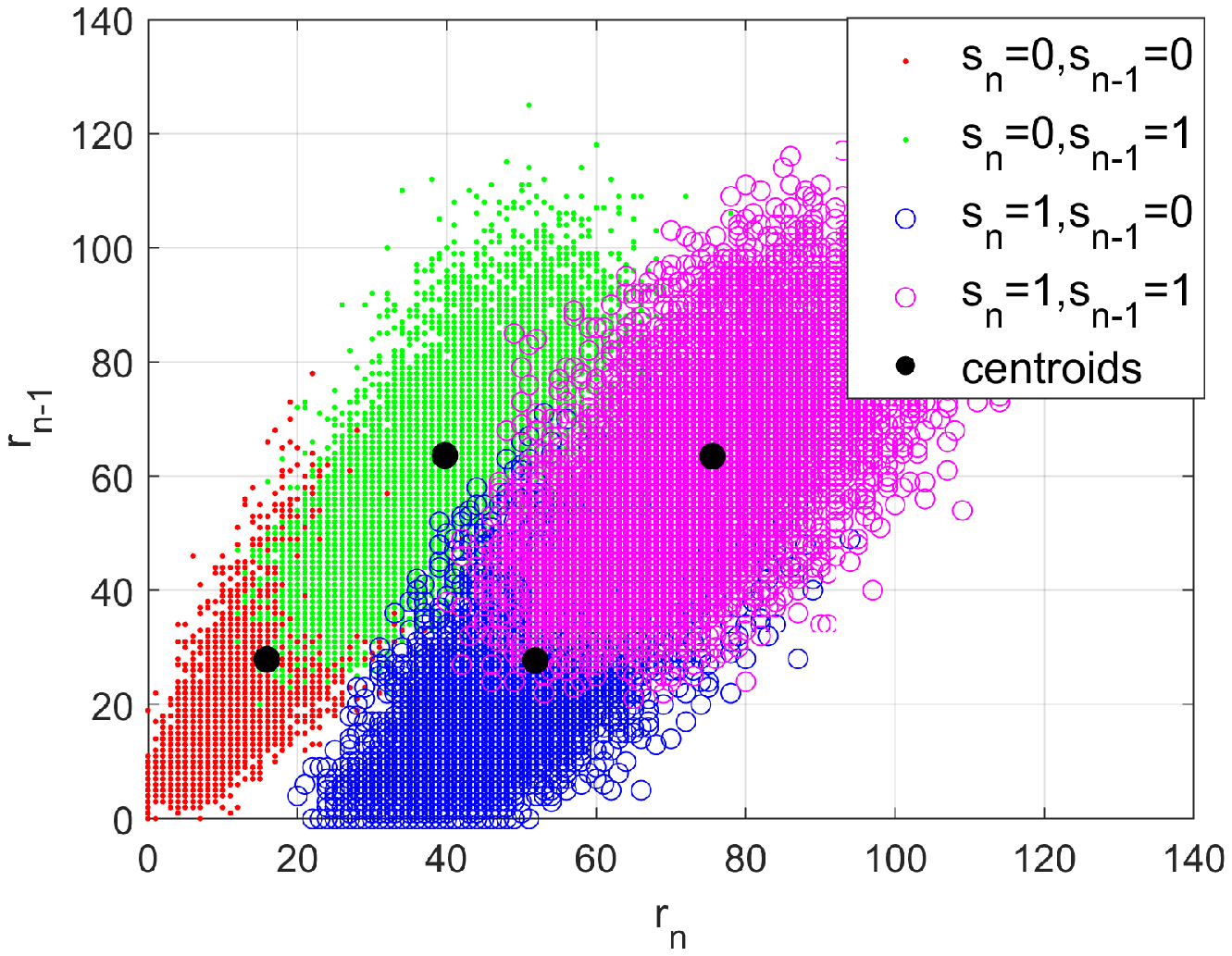}}
\subfigure[Direct clustering-based inference]{\label{fig:simple_cluster_2D_w_tau_short_a} \includegraphics[width=0.45\textwidth]{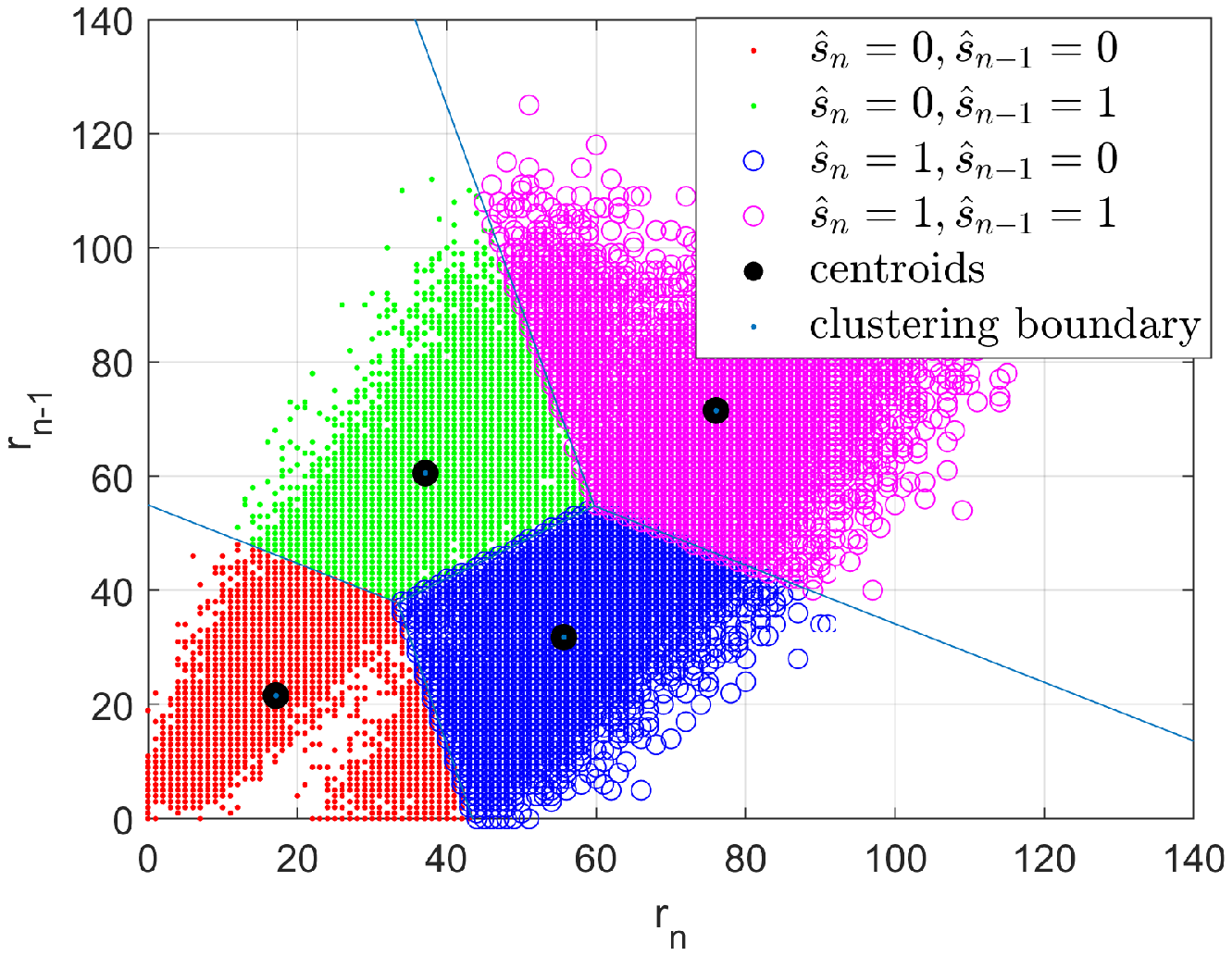}}
\subfigure[Clustering-plus-threshold detection ]{\label{fig:simple_cluster_2D_w_tau_short_b} \includegraphics[width=0.45\textwidth]{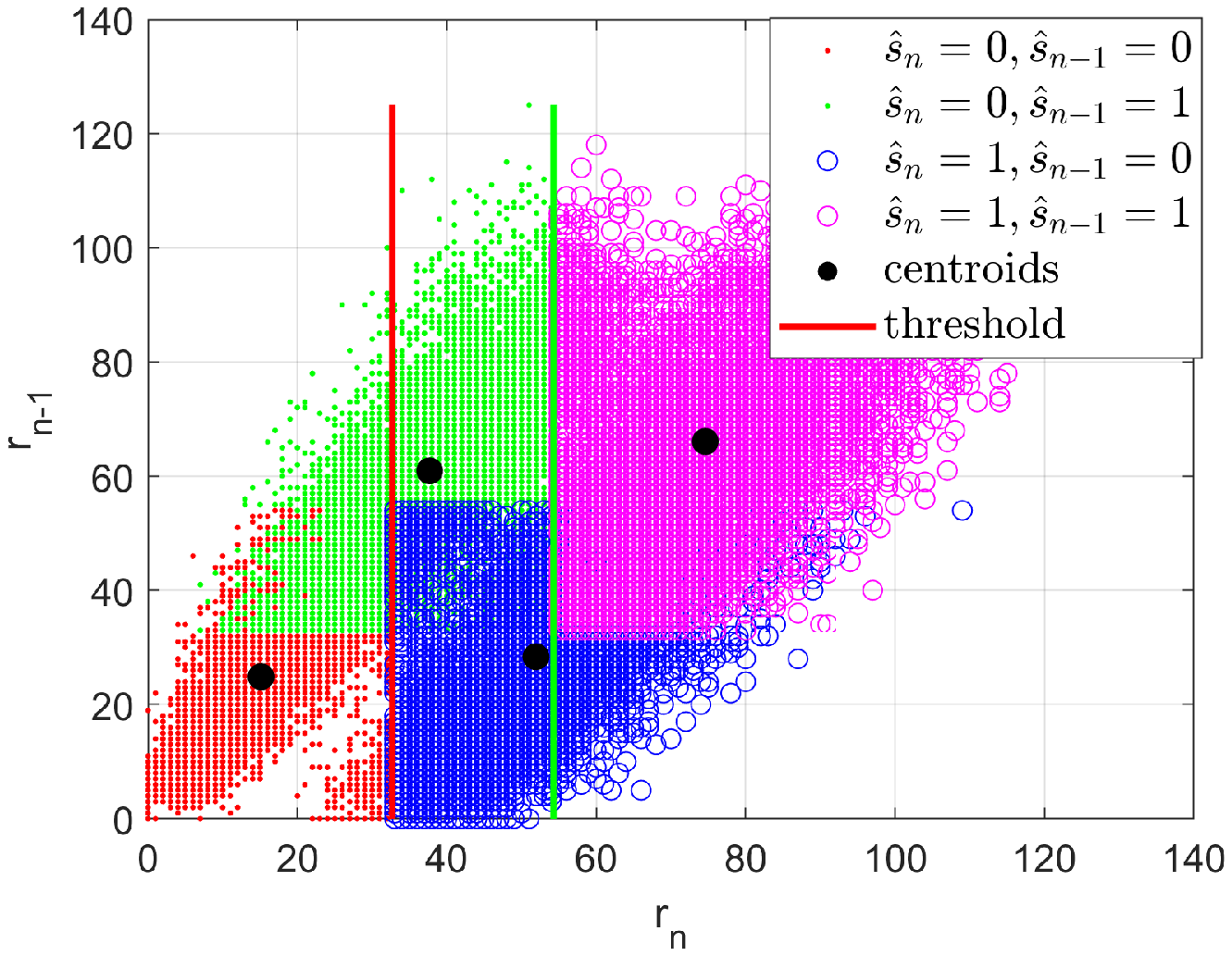}}
\caption{Distribution of $ [r_n,r_{n-1}]|_{\hat{s}_n,\hat{s}_{n-1}} $, estimated centroids and clustering boundaries/thresholds  ($ \mathcal{L}=1 $, $T=20\ \Delta T$).}
\label{fig:simple_cluster_2D_w_tau_short}
\end{figure}

\begin{figure}[!t]
\centering
\subfigure[Direct clustering-based inference]{\label{fig:simple_cluster_5D_w_tau_short_a} \includegraphics[width=0.45\textwidth]{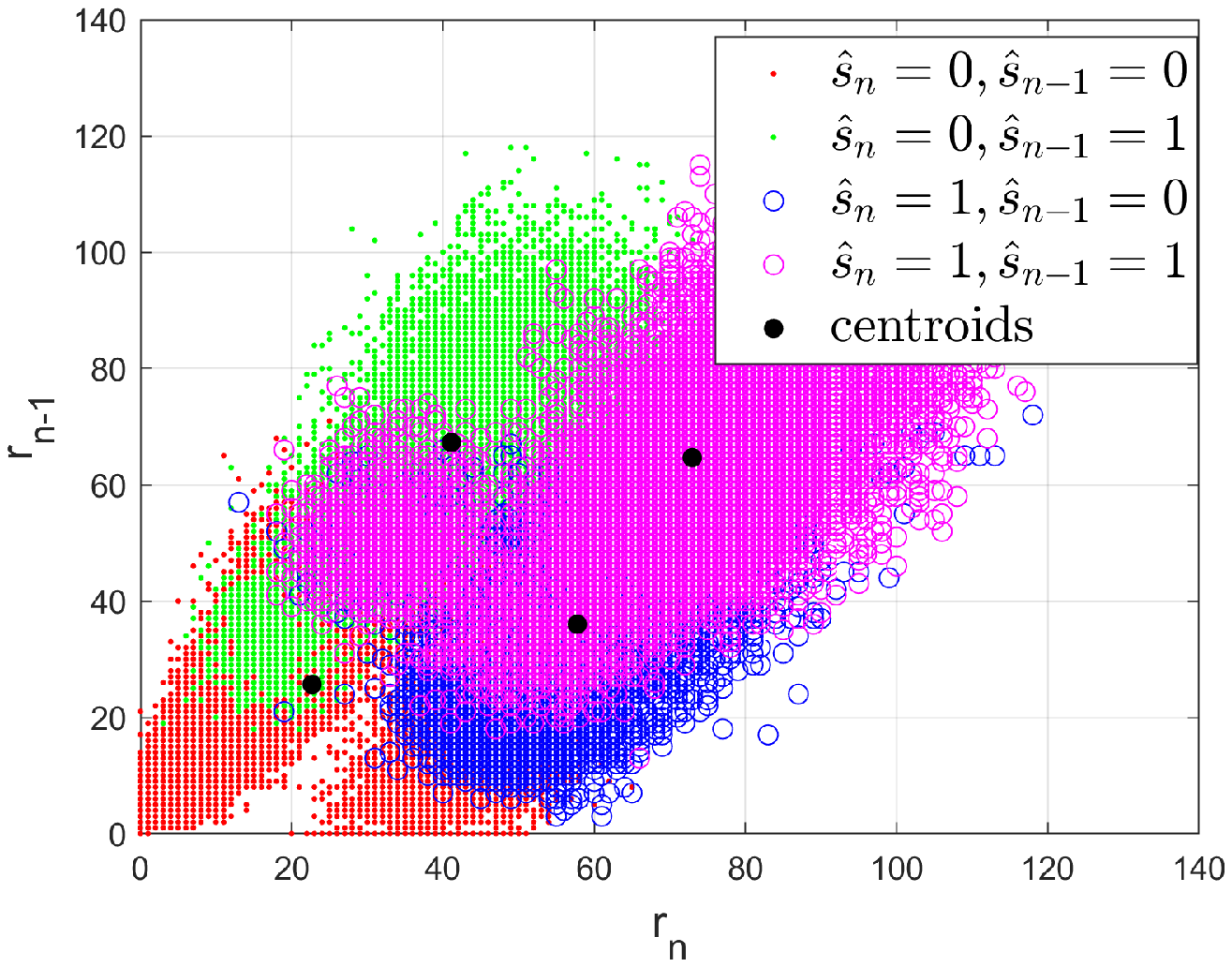}}
\subfigure[Clustering-plus-threshold detection]{\label{fig:simple_cluster_5D_w_tau_short_b} \includegraphics[width=0.45\textwidth]{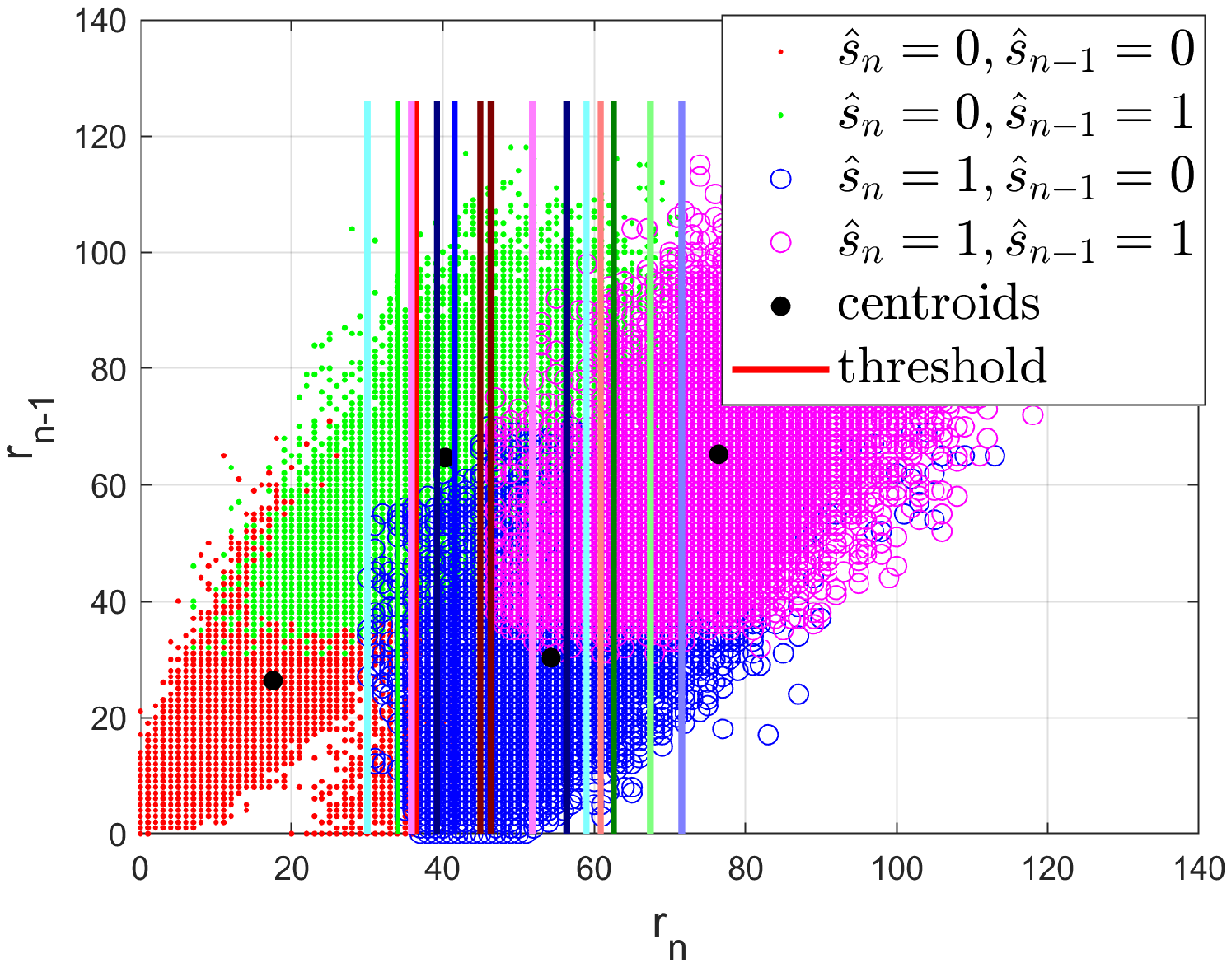}}
\caption{ Distribution of $ [r_n,r_{n-1}]|_{\hat{s}_n,\hat{s}_{n-1}} $ and estimated centroids ($ \mathcal{L}=4 $, $T=20\ \Delta T$). } 
\label{fig:simple_cluster_5D_w_tau_short}
\end{figure}

In order to appreciate the difference between the two algorithms and the advantages and limitations of clustering and using or not using the detection thresholds in (\ref{equ:threshold_reformulated}), we provide some results in Figs. \ref{fig:simple_cluster_2D_w_tau}-\ref{fig:simple_cluster_5D_w_tau_short}. These figures are obtained as follows. Consider, for example, the case study $\mathcal{L}=1$ for ease of exposition. After estimating the received data $\hat{s}_n$, all the pairs of observations $[r_n, r_{n-1}]$ that are detected as $[\hat{s}_n, \hat{s}_{n-1}] = [0,0]$, $[\hat{s}_n, \hat{s}_{n-1}] = [0,1]$, $[\hat{s}_n, \hat{s}_{n-1}] = [1,0]$, and $[\hat{s}_n, \hat{s}_{n-1}] = [1,1]$ are depicted in red, green, blue, and magenta colors, respectively. The estimated centroids are depicted as black dots. The figures report the empirical decision boundaries that are obtained after clustering (Algorithm \ref{alg:clustering_based_inference}) and after applying the empirical thresholds estimated from the centroids (Algorithm \ref{alg:clustering_plus_threshold}). We note that the two algorithms provide different results, and their advantages and limitations depend on the severity of the ISI and number of memory bits. It is worth mentioning that in Figs. \ref{fig:simple_cluster_5D_w_tau} and \ref{fig:simple_cluster_5D_w_tau_short}, due to the many memory bits used, there exist more than four clusters and the dimension of the clusters is greater than two. To be able to illustrate the results, we merge together the clusters whose $\mathcal{L}$-tuples $[r_n, r_{n-1}, \ldots, r_{n-\mathcal{L}}]$ have the same pair $[\hat s_{n}, \hat s_{n-1}]$. By using this approach, we obtain again four clusters that can be readily represented. The merged clusters are visualized by using the same color-based code as for $\mathcal{L}=1$.

By direct inspection of the figures, the following conclusions can be drawn. From Fig. \ref{fig:simple_cluster_2D_w_tau}, i.e., for mild ISI, we observe that Algorithm \ref{alg:clustering_based_inference} outperforms Algorithm \ref{alg:clustering_plus_threshold} if $\mathcal{L}=1$. In particular, Fig. \ref{fig:simple_cluster_2D_w_tau_a} yields decision regions that are closer, as compared with Fig. 5(c), to the theoretical ones in Fig. \ref{fig:simple_cluster_2D_w_tau_real}. From Fig. \ref{fig:simple_cluster_5D_w_tau}, on the other hand, we obtain an opposite trend if $\mathcal{L}=4$. Therefore, we evince that Algorithm \ref{alg:clustering_based_inference} with few memory bits may be a sufficiently good approach for non-coherent data detection under mild ISI. Under severe ISI, on the other hand, the performance trends are different, as illustrated in Figs. \ref{fig:simple_cluster_2D_w_tau_short} and \ref{fig:simple_cluster_5D_w_tau_short}. We observe, in particular, that the best estimation performance is obtained by using Algorithm \ref{alg:clustering_plus_threshold} with a large number of memory bits, i.e., $\mathcal{L}=4$ in Fig. \ref{fig:simple_cluster_5D_w_tau_short_b}. Errors are still clearly visible, but the estimated clusters of points are closer to the expected ones, i.e., those shown in Fig. \ref{fig:simple_cluster_2D_w_tau_real_short}. It is worth mentioning that in the presence of severe ISI the combination of clustering methods and detection thresholds provides better performance than using only clustering. The use of detection thresholds, in particular, allows us to correct the increased number of mis-detections that originate from increasing the dimension of clustering. These \textit{qualitative} conclusions that are drawn from the direct inspection of the estimated clusters from the detected bits are corroborated in Section \ref{sec:results_analysis} with the aid of BER simulations.

\section{Iterative Algorithm for Computing the Initial Centroids} \label{sec:non_coherent_approaches_iterative}
In the previous section, we have illustrated the core ideas of the proposed clustering-based algorithms for non-coherent detection in MC systems, which may or may not use detection thresholds. The comparison of, e.g., Fig. \ref{fig:simple_cluster_2D_w_tau_real_short} and Fig. \ref{fig:simple_cluster_5D_w_tau_short_b} reveals, however, that detection errors still exist, especially in the presence of non-negligible ISI. The sources of error include the initial estimates of the centroids and the  resulting estimates of the detection thresholds. In this section, motivated by these considerations, we introduce improved, iterative-based, clustering-based methods in order to enhance the detection performance in the presence of severe ISI.

The initial centroids obtained from $ \max(\bm{r}) $, e.g., $[0,0]$,  $ [0,\max(\bm{r})]$,  $ [\max(\bm{r}),0] $ and $ [\max(\bm{r}), $ $ \max(\bm{r})]  $ if $\mathcal{L}=1$, are, in general, not always sufficiently close to the theoretical centroids. This may result in detection errors. In this section, therefore, we introduce a more accurate iterative-based approach for estimating the initial centroids.

In order to illustrate the proposed approach, let us consider $\mathcal{L}=1$ (two-dimensional clustering) and the cluster whose label is $ [s_i,s_{i-1}]=[0,0] $.  Based on (\ref{equ:2D_prac_centers}), the theoretical centroid that corresponds to $ [s_i,s_{i-1}]=[0,0] $ is $[\bar{r}_i|_{s_i=0,s_{i-1}=0}, \bar{r}_{i-1}|_{s_i=0,s_{i-1}=0}]$, where: 
\begin{equation}
    \bar{r}_{i-1}|_{s_i=0,s_{i-1}=0} = \bar{r}_{i-1}|_{s_{i-1}=0} =  \sum_{j=1}^LC_j/2 + \bar{\lambda}_0T \label{equ:example_iter_r_i1}
\end{equation}
\begin{equation}
    \bar{r}_i|_{s_i=0,s_{i-1}=0} =  \sum_{j=2}^LC_j/2 + \bar{\lambda}_0T \label{equ:example_iter_r_i}
\end{equation}

From (\ref{equ:example_iter_r_i1}) and (\ref{equ:example_iter_r_i}), we observe that a good estimate for the initial centroid is $[a,a]$ with $ a=\sum_{j=1}^LC_j/2 + \bar{\lambda}_0T $, since this estimate would be close to the theoretical centroid, and, in general, it is closer than the initial centroid $[0,0]$ that was used in the previous section. Even though $a$ depends on $C_j$ for $j=1, 2, \ldots, L$, an estimate for it can be obtained from one-dimensional clustering without any prior CSI. In particular, $a$ is, by definition, approximately equal to $a = \bar{r}_i|_{s_i=0} \simeq \hat{\bar{r}}_i|_{s_i=0}$, where $\hat{\bar{r}}_i|_{s_i=0}$ is the estimated centroid obtained from one-dimensional clustering. A similar approach can be used to estimate the centroids of the other clusters. {\color{black} The new initial centroids constructed by using $\hat{\bar{r}}_i|_{s_i}$ are depicted in Fig. \ref{fig:kmeans_iter_updated_centroids} for completeness. By direct inspection, we can infer that they are closer to the theoretical centroids, and, therefore, are expected to provide better performance.}

In general, the proposed approach lies in setting the initial centroids for $\mathcal{L}$-dimensional clustering from the estimated centroids obtained by applying $(\mathcal{L}-1)$-dimensional clustering. In turn, the initial centroids for $(\mathcal{L}-1)$-dimensional clustering are obtained from the estimated centroids obtained by aplying $(\mathcal{L}-2)$-dimensional clustering. This procedure can be iterated until one-dimensional clustering, whose two initial centroids can be initialized to $0$ and $ \max(\bm{r}) $.

The proposed approach for setting the initial centroids in an iterative-based manner is summarized in Algorithm \ref{alg:iterative_clustering_based_inference} and Algorithm \ref{alg:iterative_clustering_plus_threshold} for application to iterative clustering-based inference and iterative clustering-plus-threshold detection, respectively. In particular, the proposed approach for setting the initial centroids works in an iterative manner, as follows: (i) one-dimensional clustering is applied to the observations $\{r_1,..., r_K\} $ by setting $ 0 $ and $  \max(\bm{r}) $ as the initial centroids; (ii) the K-means clustering algorithm is applied and the estimated centroids $ \hat{\bar{r}}_i|_{s_i=0} $ and $ \hat{\bar{r}}_i|_{s_i=1} $ are obtained; (iii) the estimated centroids $ \hat{\bar{r}}_i|_{s_i=0} $ and $ \hat{\bar{r}}_i|_{s_i=1} $ are used to construct new initial centroids for application to two-dimensional clustering. In particular, the four initial centroids are set to $ [\hat{\bar{r}}_i|_{s_i=0}, \hat{\bar{r}}_i|_{s_i=0}] $, $ [\hat{\bar{r}}_i|_{s_i=0}, \hat{\bar{r}}_i|_{s_i=1}] $,  $ [\hat{\bar{r}}_i|_{s_i=1}, \hat{\bar{r}}_i|_{s_i=0}] $, and  $ [\hat{\bar{r}}_i|_{s_i=1}, \hat{\bar{r}}_i|_{s_i=1}] $; (iv) two-dimensional clustering is applied to the vectors of observations $ \{[r_2,r_1],..., [r_K,r_{K-1}]\} $ by using the initial estimated centroids; (v) the K-means (two-dimensional) clustering algorithm is applied again and the estimated two-dimensional centroids  $ [\hat{\bar{r}}_i,\hat{\bar{r}}_{i-1}]|_{s_i,s_{i-1}}$ are obtained; (vi) the procedure is iterated until the clustering dimension $\mathcal{L}+1$. 

In general terms, the initial centroids for application to the $(l+1)$-dimensional clustering can be constructed from the estimated centroids obtained from $l$-dimensional clustering. In mathematical terms, let us denote the estimated centroids from $l$-dimensional clustering by $ [\hat{\bar{r}}_m, ..., \hat{\bar{r}}_{m-l+1}]|_{s_{m-j},0\leq j\leq l-1} $. The initial centroid $ \bm{\mu}_k $ that corresponds to the label $ [s_i, ..., s_{i-l+1},$  $ s_{i-l}] $ for $(l+1)$-dimensional clustering can be obtained as follows:
\begin{equation}
\bm{\mu}_k = [\hat{\bar{r}}_m|_{s_{m-j}=s_{i-j},0\leq j\leq l-1}, \hat{\bar{r}}_m|_{s_{m-j}=s_{i-1-j},0\leq j\leq l-1}, \hat{\bar{r}}_{m-1}|_{s_{m-j}=s_{i-1-j},1\leq j\leq l-1},...,\hat{\bar{r}}_{m-l+1}|_{s_{m-l+1}=s_{i-l}}] \label{equ:iter_centroid}
\end{equation}
which does not require any prior CSI information.

\begin{figure*}[htb]
\centering
\begin{minipage}[t]{.75\textwidth}
\begin{algorithm}[H]
 \caption{{Iterative clustering-based inference}}
 \begin{algorithmic}[1]
 \STATE Set the initial centroids $ \bm{\mu}_0 = 0 $ and $ \bm{\mu}_1 = \max(\bm{r}) $
 \FOR{ $ l=1 $ to $ \mathcal{L}+1 $ }
 \STATE Construct the data $ \bm{r}_n = [r_n,...,r_{n-l+1}]  $
 \STATE Cluster $ \bm{r}_n  $ using the K-means algorithm with the initial centroids $ \bm{\mu}_k $ 
 \STATE Set the new initial centroids to $ \bm{\mu}_k $ by using (\ref{equ:iter_centroid})
 \ENDFOR
  \STATE Infer $ \hat{s}_n $ from the indicator variables $ \kappa_{n,k} $
 \end{algorithmic} 
 \label{alg:iterative_clustering_based_inference}
\end{algorithm}
\end{minipage}
\begin{minipage}[t]{.75\textwidth}
\begin{algorithm}[H]
 \caption{{Iterative clustering-plus-threshold detection}}
 \begin{algorithmic}[1]
 \STATE Set the initial centroids $ \bm{\mu}_0 = 0 $ and $ \bm{\mu}_1 = \max(\bm{r}) $
 \FOR{ $ l=1 $ to $ \mathcal{L}+1 $ }
 \STATE Construct the data $ \bm{r}_n = [r_n,...,r_{n-l+1}]  $
 \STATE Cluster $ \bm{r}_n  $ using the K-means algorithm with the initial centroids $ \bm{\mu}_k $ 
 \STATE Set the new initial centroids to $ \bm{\mu}_k $ by using (\ref{equ:iter_centroid})
 \ENDFOR
  \STATE Obtain $ \hat{\bar{r}}_i|_{\hat{s}_{i-j},0\leq j\leq \mathcal{L}} $ from the estimated centroids $ \hat{\bm{\mu}}_k $
  \STATE Compute the detection thresholds using (\ref{equ:threshold_reformulated})
  \STATE Detect the symbols using (\ref{equ:detection_with_threshold}) 
 \end{algorithmic} 
 \label{alg:iterative_clustering_plus_threshold}
\end{algorithm}
\end{minipage}
\end{figure*}

\begin{figure}[!t]
\centering
\subfigure[Iterative clustering-based inference]{\label{fig:iterative_cluster_5D_w_tau_short_a} \includegraphics[width=0.45\textwidth]{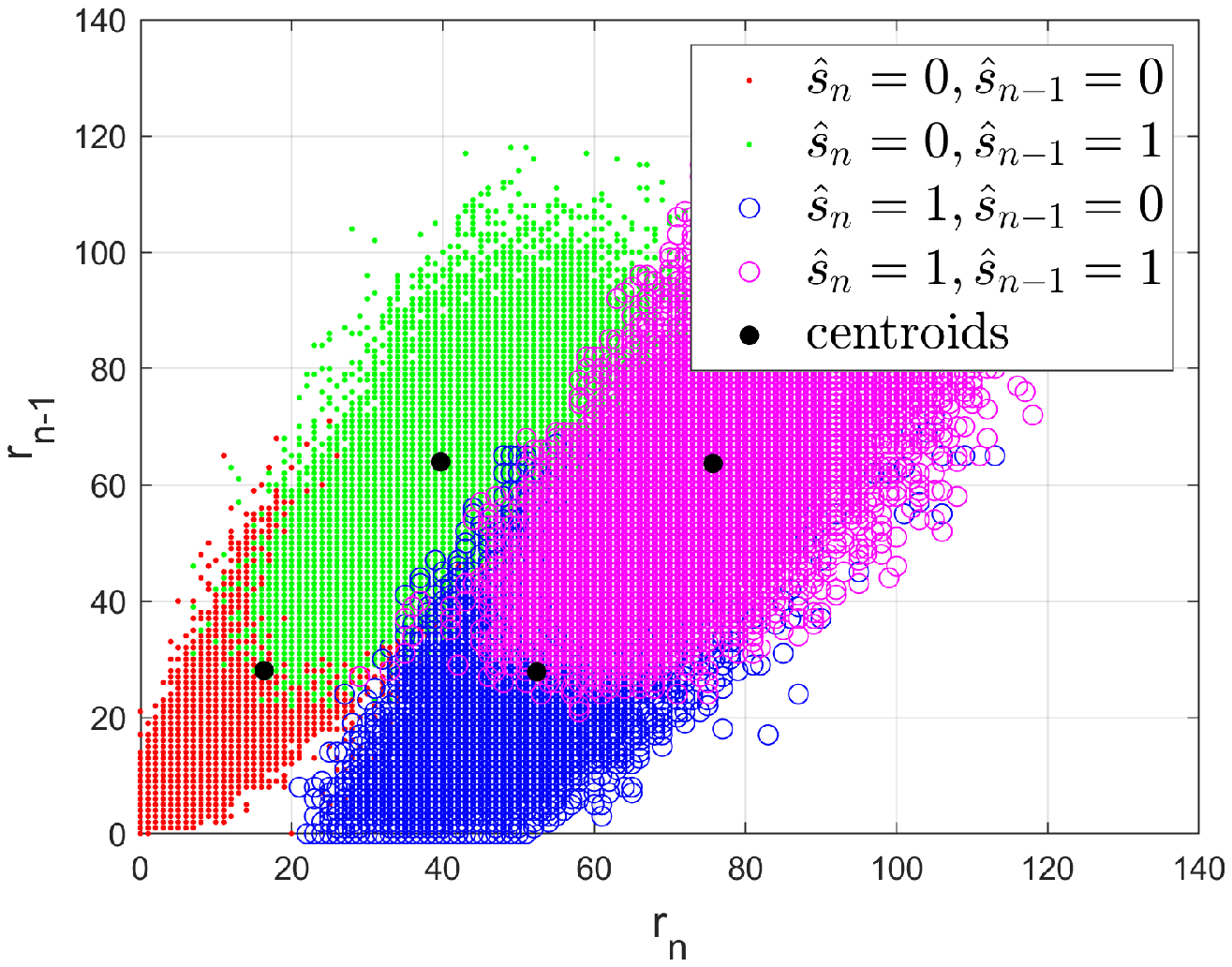}}
\subfigure[Iterative clustering-plus-threshold detection ]{\label{fig:iterative_cluster_5D_w_tau_short_b} \includegraphics[width=0.45\textwidth]{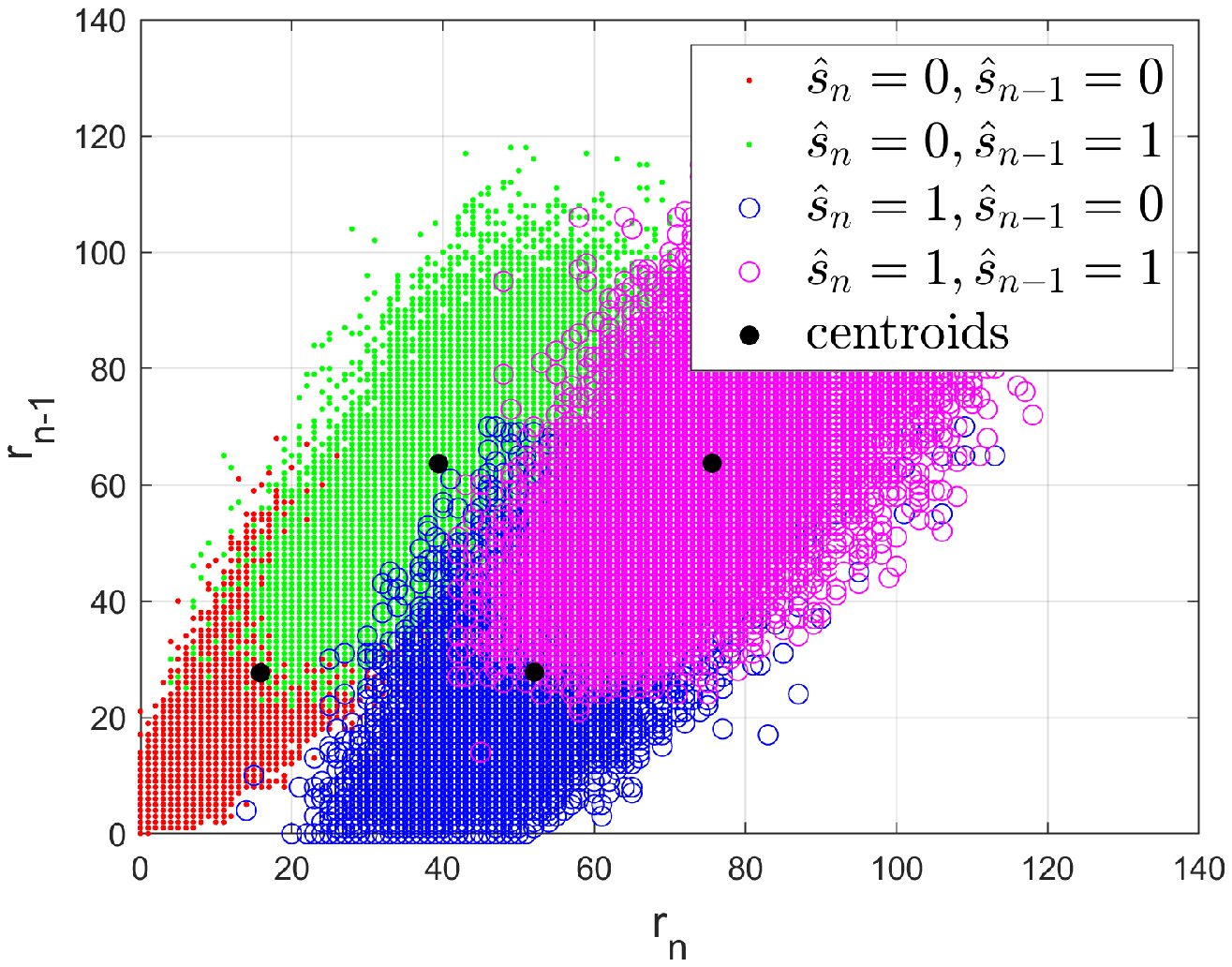}}
\caption{Distribution of $ [r_n,r_{n-1}]|_{\hat{s}_n,\hat{s}_{n-1}} $ and estimated centroids ($ \mathcal{L}=4 $, $T = 20 \Delta T$). } 
\label{fig:iterative_cluster_5D_w_tau_short}
\end{figure}

By considering the same case study as in Fig. \ref{fig:simple_cluster_5D_w_tau_short}, we illustrate the performance obtained by employing the proposed iterative approach in Algorithms \ref{alg:iterative_clustering_based_inference} and \ref{alg:iterative_clustering_plus_threshold} in Fig. \ref{fig:iterative_cluster_5D_w_tau_short}. We observe that, even in the presence of severe ISI, much better estimation performance is obtained by a more accurate initial estimate of the initial centroids. These results confirm the effectiveness of the proposed iterative-based approach.

\section{Numerical Results }\label{sec:results_analysis}

In this section, we report some simulation results in order to quantitatively analyze the performance of the proposed non-coherent detectors. {\color{black} In particular, we compare the proposed methods with a few benchmark schemes available in the literature}. The simulation parameters are listed in Table \ref{Tab:parameters_short_slot}. 

\subsection{Computational Complexity}
Before analyzing the error performance of the proposed clustering-based algorithms, we study the computational complexity of the proposed schemes in Table \ref{Tab:Algorithm_comparison}. In our study, we assume that Step I and Step II of the K-means clustering algorithms are repeated $ P $ times.  For instance, if we consider $K$ received signals stored and $\mathcal{L}$ memory bits, (the centroid dimension is $\mathcal{L}+1$ and there are  $2^{\mathcal{L}+1}$ initial centroids), in Step I of the K-means clustering algorithm, the evaluation of the distance between one point and a centroid requires $2\mathcal{L}+1$ additions and $\mathcal{L}+1$ multiplications. For all points, thus, we obtain $2^{\mathcal{L}+1}K (2\mathcal{L}+1)$ additions and $2^{\mathcal{L}+1}K(\mathcal{L}+1)$ multiplications. In Step II, we need $(K-2^{\mathcal{L}+1})(\mathcal{L}+1)$ additions and $2^{\mathcal{L}+1}$ multiplications. Since the computational complexity of Step II is negligible compared with the computational complexity of Step I, we omit the complexity of Step II and report only the computational complexity of Setp I in Table \ref{Tab:Algorithm_comparison}.

\begin{table}[!t] 
\centering
\caption{{Simulation parameters}}
\label{Tab:parameters_short_slot}
\begin{tabular}{l|l}
\hline
{Parameter} & {Value} \\ \hline
{$ \bar{\lambda}_0 $}  & {$ 100 s^{-1} $} \\ 
{Receiver radius $ r $} & {45 nm} \\ 
{Distance $ d $} & {500 nm}  \\ 
{Diffusion coefficient $ D $} & {$4.265\times 10^{-10} m^2/s$} \\ 
{Discrete time length  $ \Delta T $}  & {9 us}  \\ 
{Channel length $ L $} & {5} \\ 
{ $ K $} & {$ 2^{12} $} \\ 
\hline
\end{tabular}
\label{tab:parameter}
\end{table}

\begin{table}[!t] 
\centering
\caption{{Computational complexity}}
\label{Tab:Algorithm_comparison}
\begin{tabular}{l|ll}
\hline
{Algorithm} & {Additions}  & {Multiplications}  \\ \hline
{Algorithm \ref{alg:clustering_based_inference},  \ref{alg:clustering_plus_threshold} } & {$2^{\mathcal{L}+1}PK(2\mathcal{L}+1)$} & {$2^{\mathcal{L}+1}PK(\mathcal{L}+1)$}  \\ 
{Algorithm \ref{alg:iterative_clustering_based_inference}, \ref{alg:iterative_clustering_plus_threshold} } & {$PK \sum\limits_{i=1}^{\mathcal{L}+1}2^{i}(2i-1)$} & {$PK\sum\limits_{i=1}^{\mathcal{L}+1}2^{i}(2i-1)$}  \\ \hline
\end{tabular}
\end{table}

\subsection{Channels in the Presence of ISI}

In order to evaluate the efficiency of the proposed schemes, we compare the BER performance of the proposed algorithms with some benchmark schemes, including: (i) a receiver that first estimates the channel by using the method proposed in \cite{jamali2016channel} by using $K_{\mathrm{train}}$ pilot symbols and then detects the bits by using the threshold in (\ref{equ:subopt_threshold_derived}); (ii) a receiver that detects the symbols by using the threshold in (\ref{equ:subopt_threshold_derived}) by assuming that the CSI is perfectly known; (iii) a sequence-based receiver \cite{ViterbiNet_ref} that detects the symbols by assuming perfect CSI (including the knowledge of $L$) and by using a multi-symbol maximum likelihood (ML) detector based on \eqref{equ:app_r_probability}; (iv) the ViterbiNet sequence-based receiver in \cite{ViterbiNet_ref}, which knows $L$ and is trained using $ 2^{10} $ pilot symbols. In order to show how the channel estimation accuracy affects the detection performance, we evaluate the benchmark scheme for $K_{\mathrm{train}}=2^{6}$ and $K_{\mathrm{train}}=2^{10}$.

\begin{figure}[!t]
\centering 
\subfigure[$ \mathcal{L}=1 $  and $ T = 30\ \Delta T $]{\label{fig:Kmeans_BER_comparison_4cluster_SL30} \includegraphics[width=0.45\textwidth]{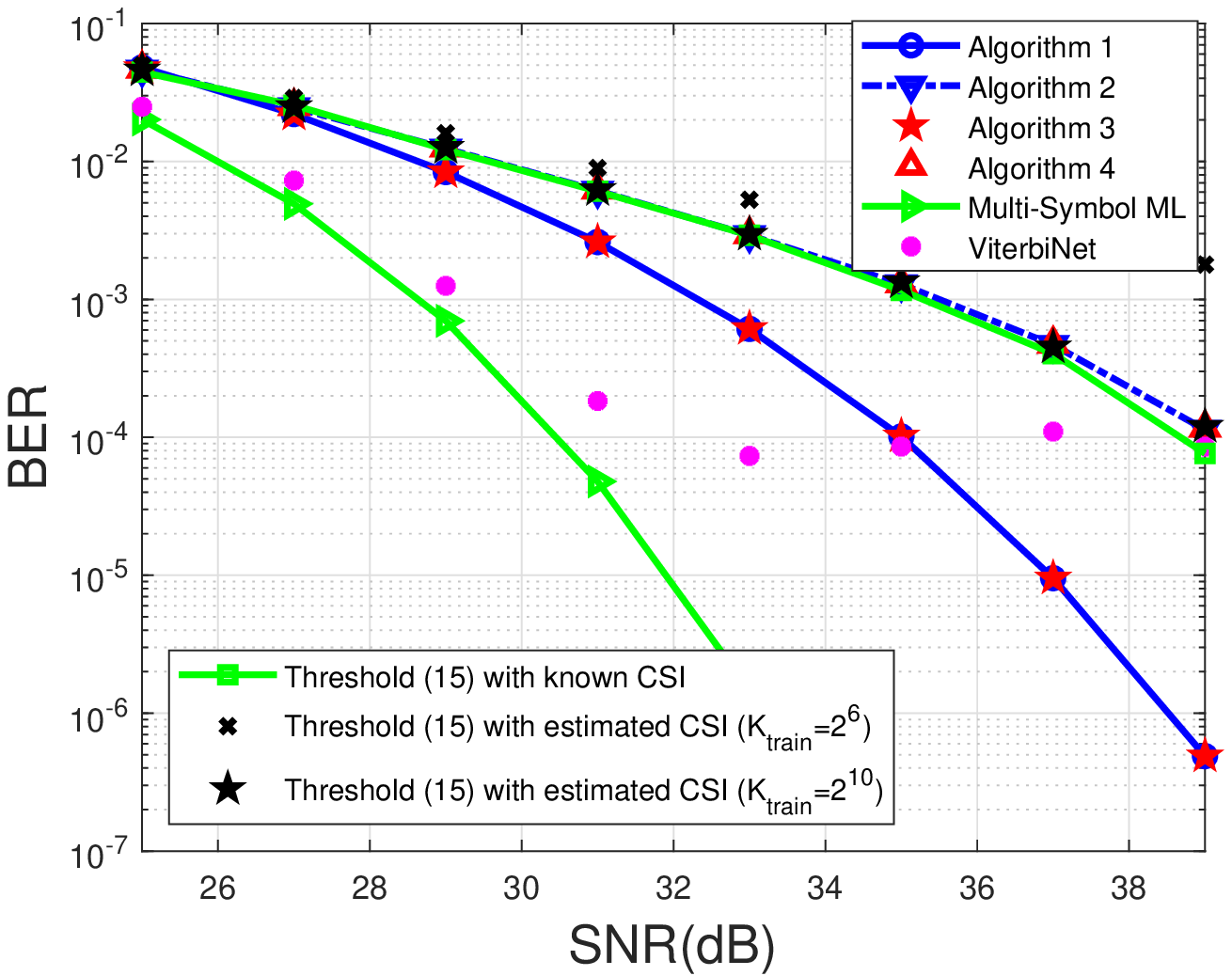}}
\subfigure[$ \mathcal{L}=4 $ and $ T = 30\ \Delta T $]{\label{fig:Kmeans_BER_comparison_32cluster_SL30} \includegraphics[width=0.45\textwidth]{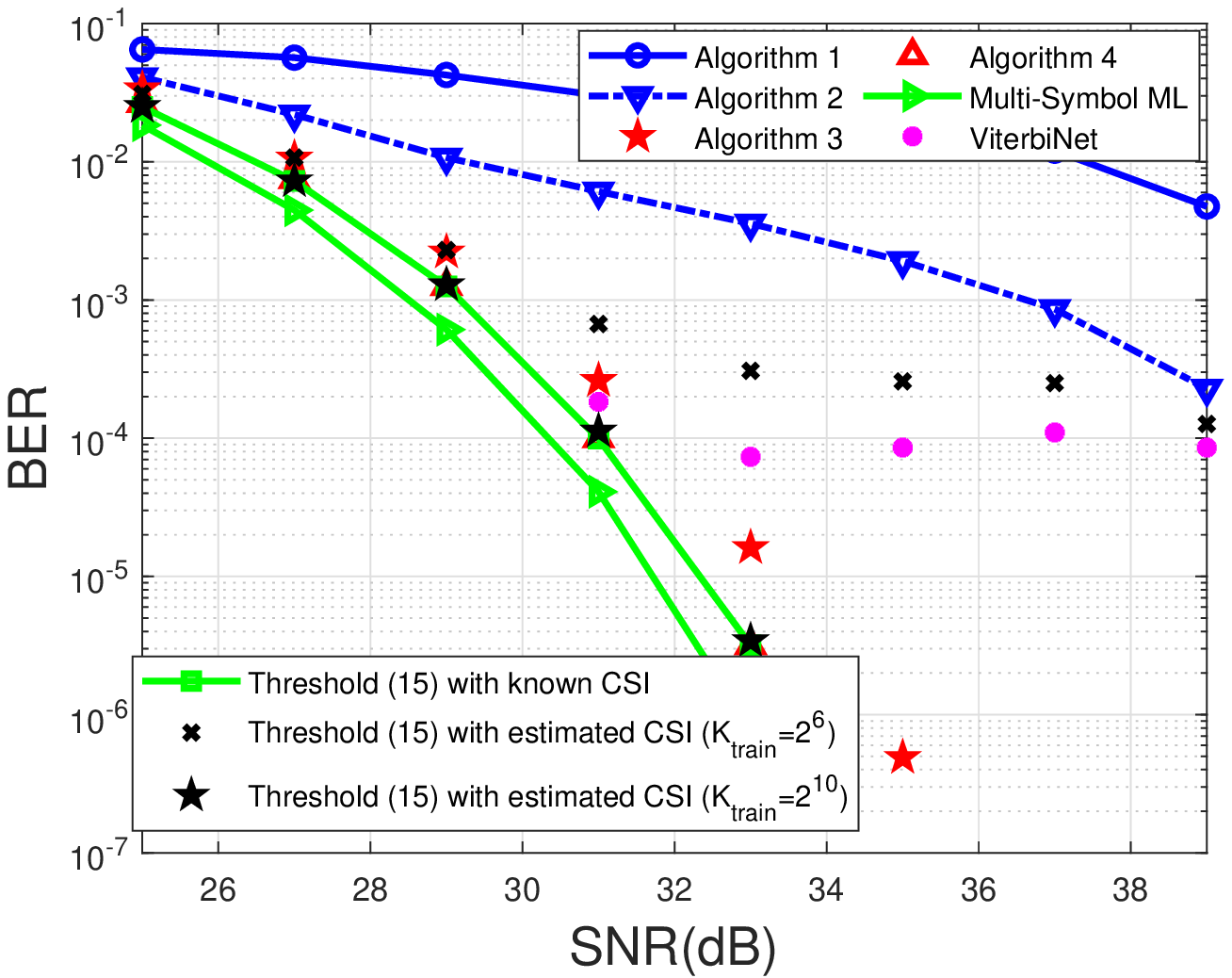}}
\subfigure[$ \mathcal{L}=1 $  and $ T = 30\ \Delta T $ (known CSI)]{\label{fig:Kmeans_BER_comparison_4cluster_SL30_withCSI} \includegraphics[width=0.45\textwidth]{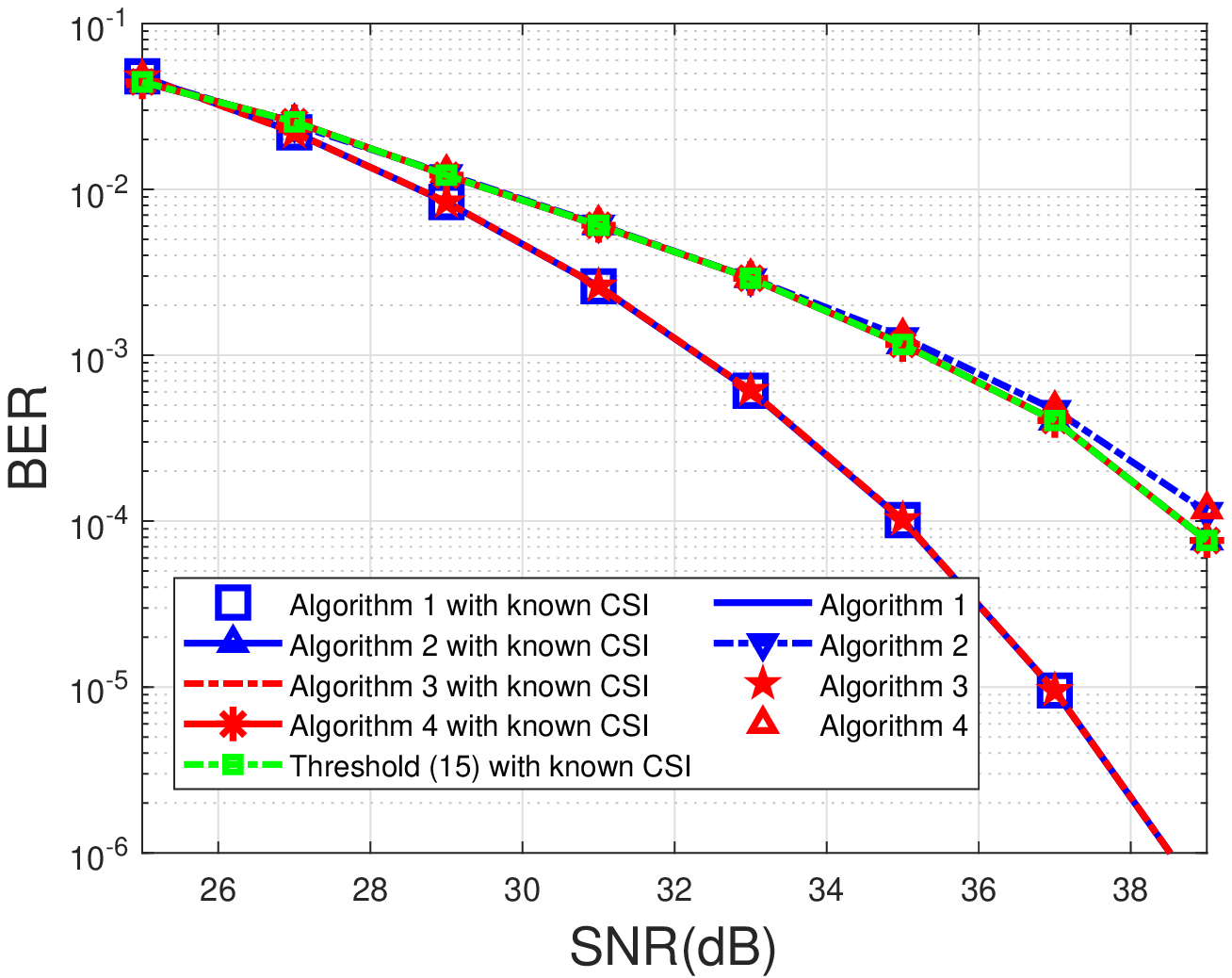}}
\subfigure[$ \mathcal{L}=4 $ and $ T = 30\ \Delta T $ (known CSI) ]{\label{fig:Kmeans_BER_comparison_32cluster_SL30_withCSI} \includegraphics[width=0.45\textwidth]{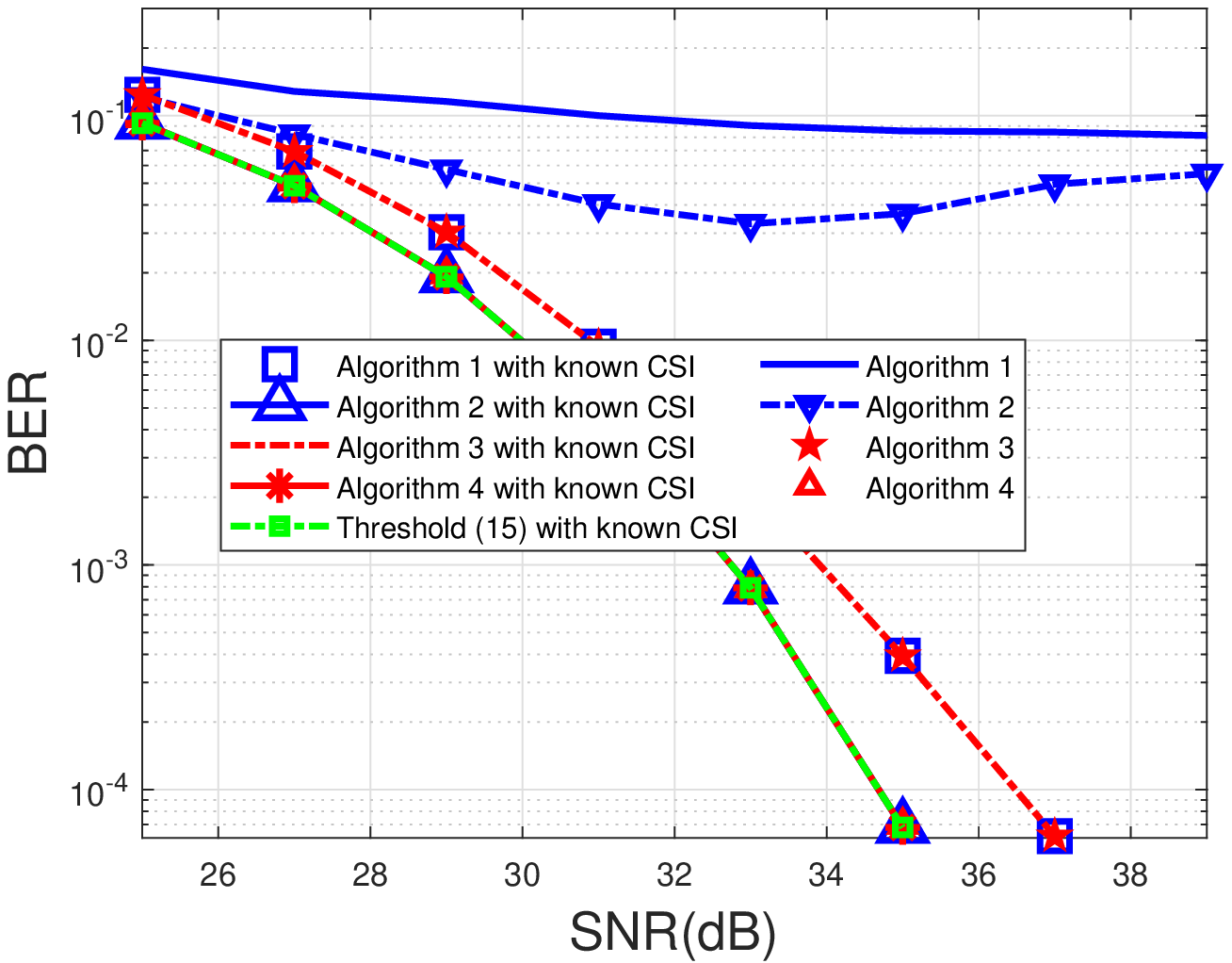}}
\caption{BER comparison of the proposed algorithms (mild ISI) }
\label{fig:Kmeans_BER_comparison_SL30}
\end{figure}

In Fig. 10 (mild ISI) and Fig. 11 (severe ISI), we compare the performance of the proposed algorithms against the considered benchmark schemes by assuming a different number of memory bits $\mathcal{L}$ and the case studies in which the CSI is unknown (practical case) and known (benchmark case). We observe that the ML-based sequence detector with perfect CSI knowledge yields the best BER performance, as expected, and provides us with a lower bound. The proposed Algorithm 3 and Algorithm 4 yield good performance if the number of memory bits is not too small, depending on the severity of the ISI. By assuming that the CSI is known (e.g., the initial centroids are perfectly known in our algorithms), we observe (see Fig. 10(c) and Fig. 10(d)) that the proposed algorithms provide good performance. This shows the robustness of the proposed approach, which works well even under known CSI. We note, in addition, that the ViterbiNet shows an error-floor, in general, because it is derived by assuming a Gaussian white noise, which is different from the considered model.

{In particular, we observe that Algorithm 1 and Algorithm 2 do not perform well, in general, in the presence of ISI because of errors in estimating the centroids. By comparing Fig. 10(c) and Fig. 10 (d), we evince that Algorithm 1 and Algorithm 2 have a worse BER if $\mathcal{L}$ increases. This is due to the high dimension of the clustered data. On the other hand, Algorithm 3 and Algorithm 4 compensate for this issue thanks to the iterative approach that is used to estimate the centroids. Based on the computational complexity that can be afforded, which depends on $\mathcal{L}$, Algorithm 3 and Algorithm 4 are the most suitable options. In particular, Fig. 10(c) shows that Algorithm 3 is the preferred choice if $\mathcal{L}=1$ and Algorithm 4 is the preferred choice if $\mathcal{L}=4$. The same trends are observed in Fig. 11. In the presence of severe ISI, it is necessary, however, to use a large number of memory bits $\mathcal{L}$ in order to outperform the ViterbiNet, as illustrated in Fig. 11(a).

}

\begin{figure}[!t]
\centering 
\subfigure[$ \mathcal{L}=1 $  and $ T = 20\ \Delta T $]{\label{fig:Kmeans_BER_comparison_4cluster_SL20} \includegraphics[width=0.45\textwidth]{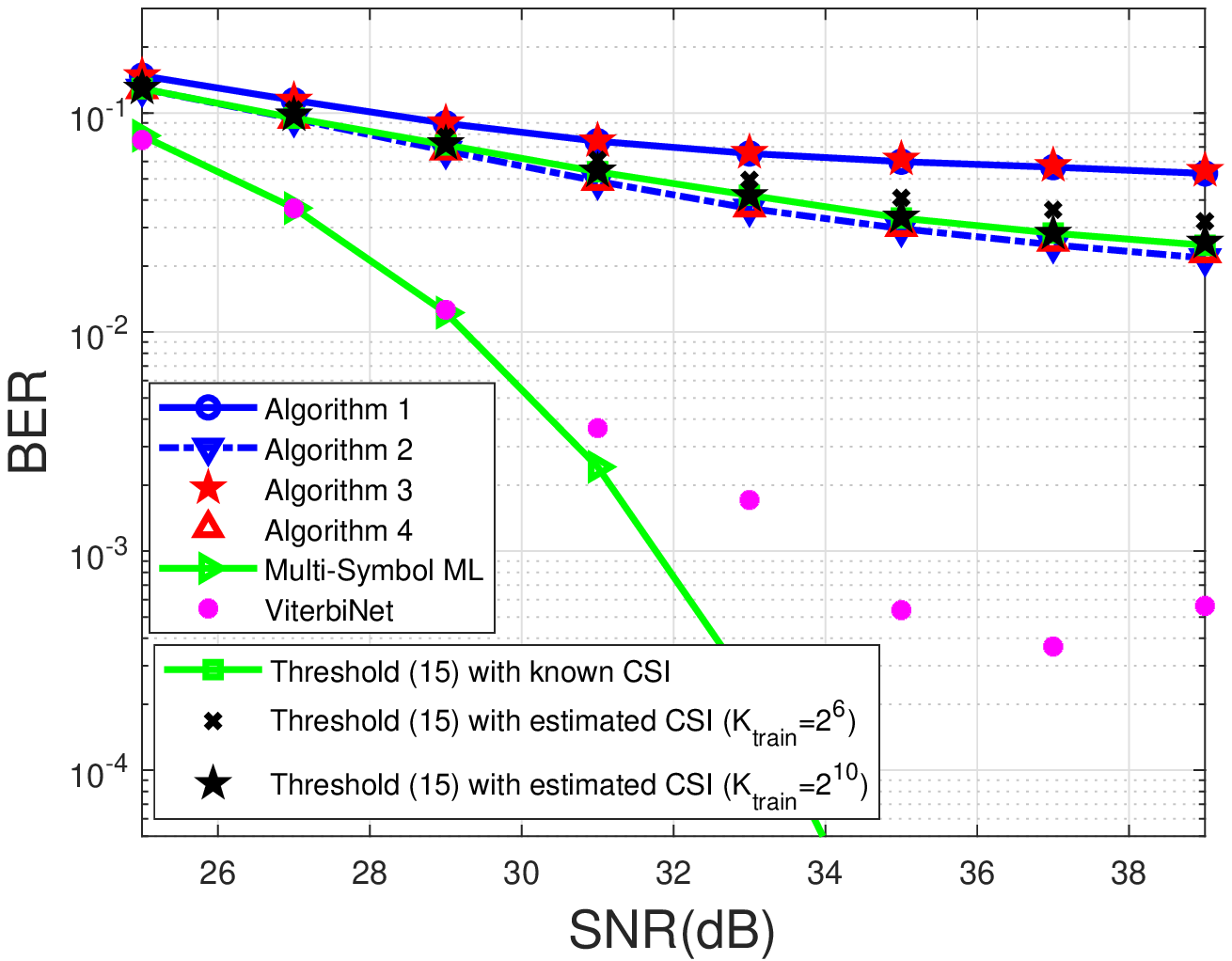}}
\subfigure[ $ \mathcal{L}=4 $ and $ T = 20\ \Delta T $]{\label{fig:Kmeans_BER_comparison_32cluster_SL20} \includegraphics[width=0.45\textwidth]{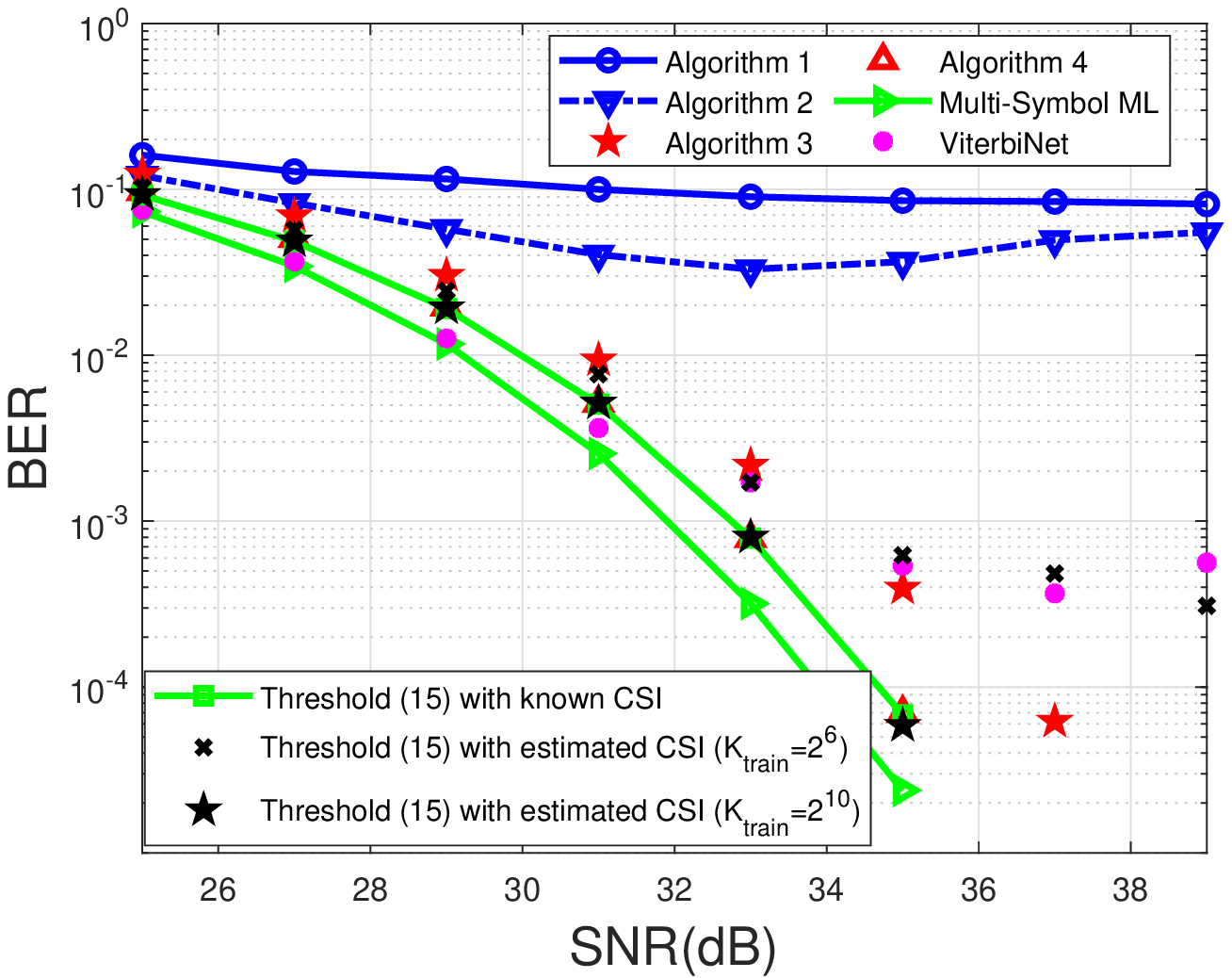}}
\caption{BER comparison of the proposed algorithms (severe ISI) }
\label{fig:Kmeans_BER_comparison_SL20}
\end{figure}


\subsection{Channels Without ISI}
In order to further assess the robustness of the proposed algorithms, we evaluate the BER performance under ISI-free channels. In particular, we compare the BER performance of our algorithms against the benchmark in \cite{Jamali_CC}, which is based on a non-coherent detector that uses constant-composition codes. In addition, we consider a benchmark that assumes that the CSI is known and uses the theoretical threshold in (\ref{equ:subopt_threshold_derived}). {The results are illustrated in Fig. 12. We observe that Algorithm 1 gives similar performance as Algorithm 3, and Algorithm 2 gives similar performance as Algorithm 4. This is because no ISI is present and the initial estimate of the centroids is almost correct. In particular, Algorithms 2 and 4 outperform Algorithms 1 and 3 because the decoding thresholds are exploited. In addition, Algorithm 2 and Algorithm 4 yield a BER that is close to the benchmark scheme with known CSI, and they outperform the benchmark in [36].}

\begin{figure}[!t]
\centering 
\includegraphics[width=0.45\textwidth]{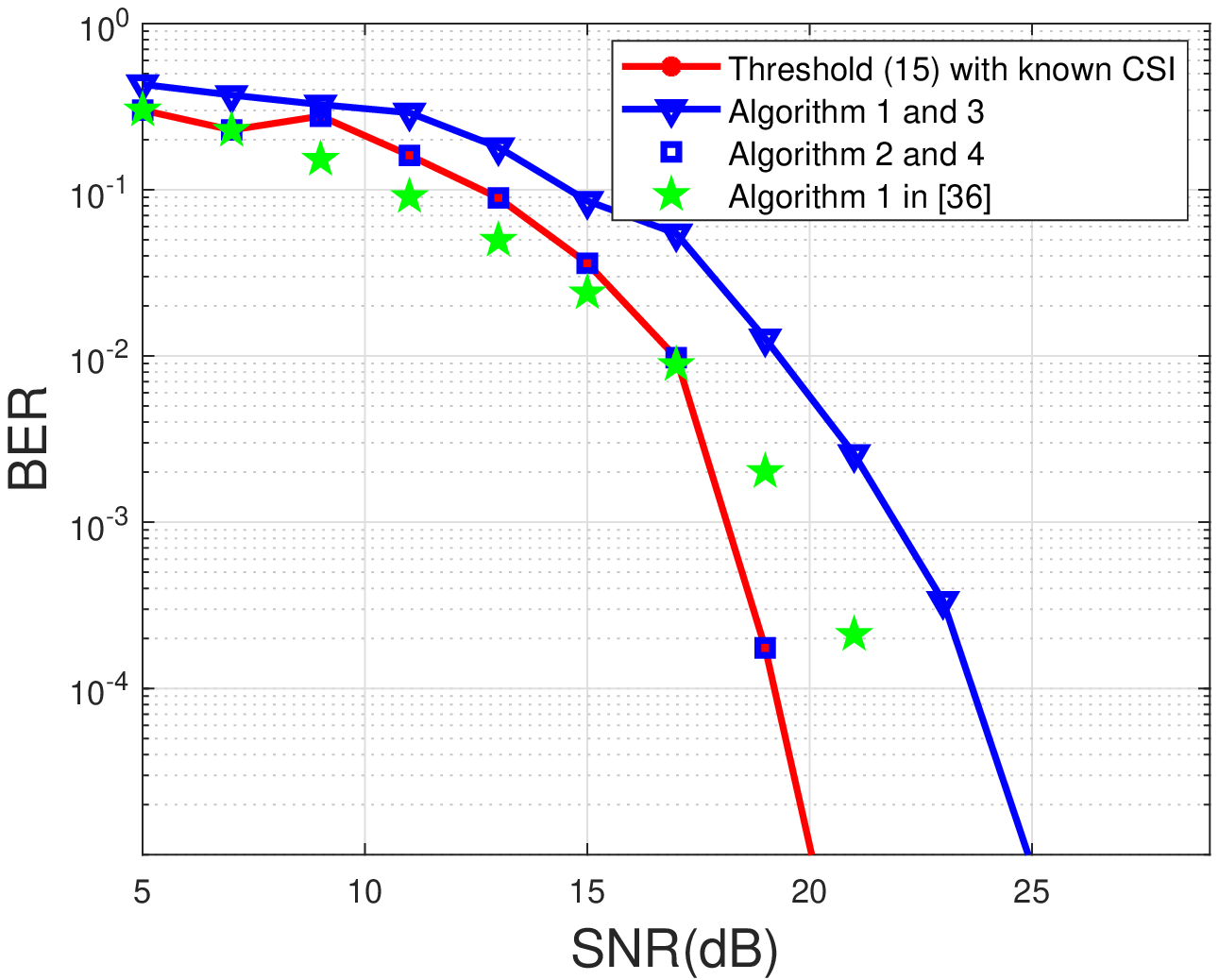}
\caption{BER comparison of the proposed algorithms (no ISI, $ T = 100\ \Delta T $ and $ \mathcal{L}=1 $) }
\label{fig:Kmeans_BER_comparison_CC_no_ISI}
\end{figure}



\subsection{Extension to Multi-Level CSK Modulation}
To explore the potential applications of the proposed clustering-based methods, we generalize the proposed approach to multi-level CSK modulation \cite{Galmes_MultiLevel}. As an example, we consider a four-level CSK modulation scheme. By using similar analytical steps as those for the derivation of \eqref{equ:subopt_threshold_derived}, e.g., for one memory symbol, we obtain the threshold:
\begin{eqnarray}
   \tau_k|_{s_{i-1}} = \frac{\bar{r}_i|_{s_i=k,s_{i-1} }-\bar{r}_i|_{s_i=k-1,s_{i-1} }}{\ln(\frac{\bar{r}_i|_{s_i=k,s_{i-1}}}{\bar{r}_i|_{s_i=k-1,s_{i-1}}} ) } 
\end{eqnarray}
by imposing $ \mathrm{Pr}_{\mathrm{approx}} (r_i|s_i=k-1,s_{i-1}) = \mathrm{Pr}_{\mathrm{approx}} (r_i|s_i=k,s_{i-1}) $. If $ r_i\leq \tau_1|_{s_{i-1}} $, then $ \hat{s}_i=0 $; if $ \tau_1|_{s_{i-1}}< r_i\leq \tau_2|_{s_{i-1}} $, then $ \hat{s}_i=1 $; if $  \tau_2|_{s_{i-1}}< r_i\leq \tau_3|_{s_{i-1}} $, then $ \hat{s}_i=2 $; otherwise, $ \hat{s}_i=3 $. 

\begin{figure}[!ht]
\centering 
\subfigure[BER comparison of the proposed algorithm (mild ISI)]{\label{fig:Kmeans_BER_MultiLevel_1} \includegraphics[width=0.45\textwidth]{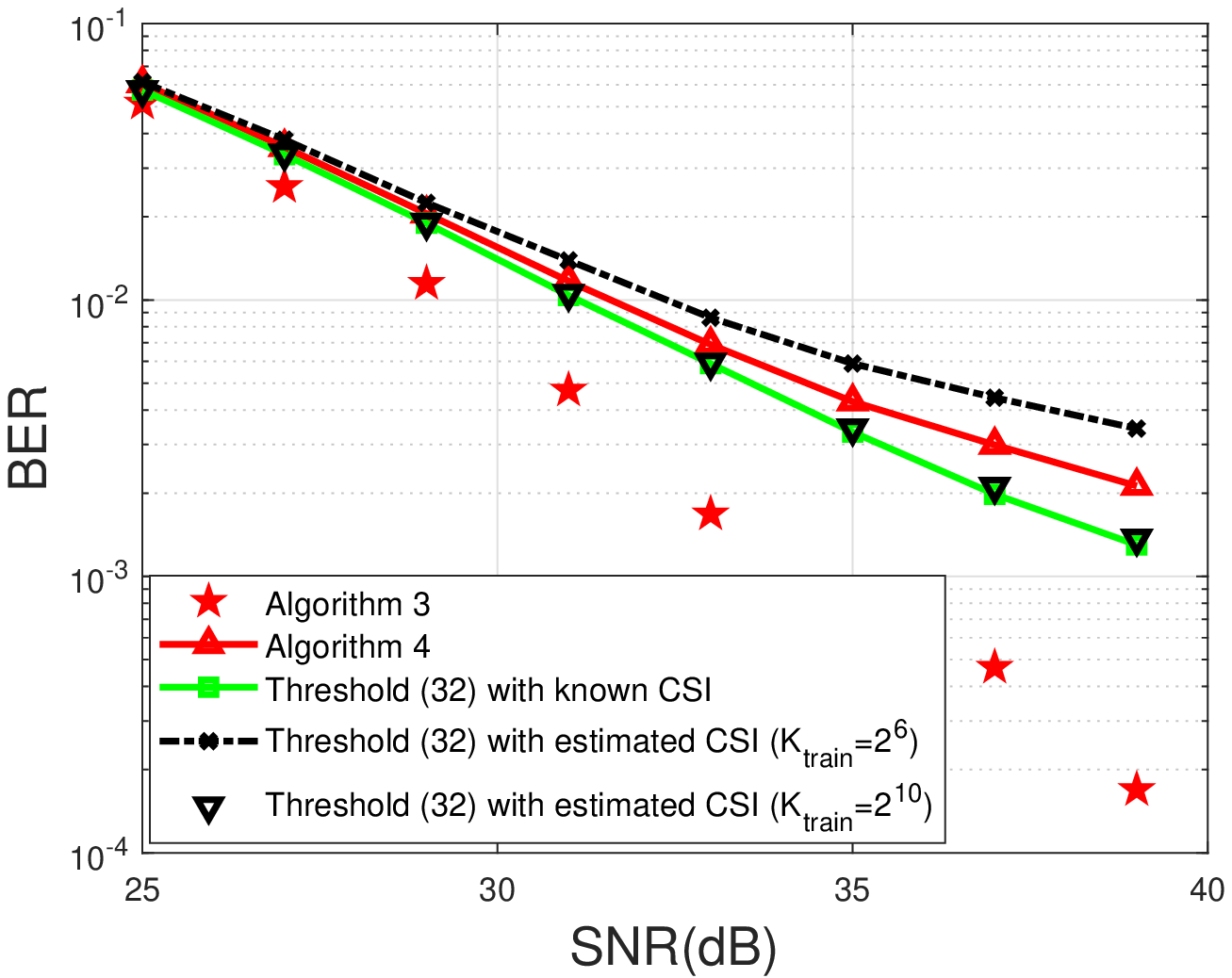}}
\subfigure[ Distribution of \texorpdfstring{$ [r_n,r_{n-1} ]|_{s_i,s_{i-1}} $}{original} with correct labels]{\label{fig:Kmeans_BER_MultiLevel_2} \includegraphics[width=0.45\textwidth]{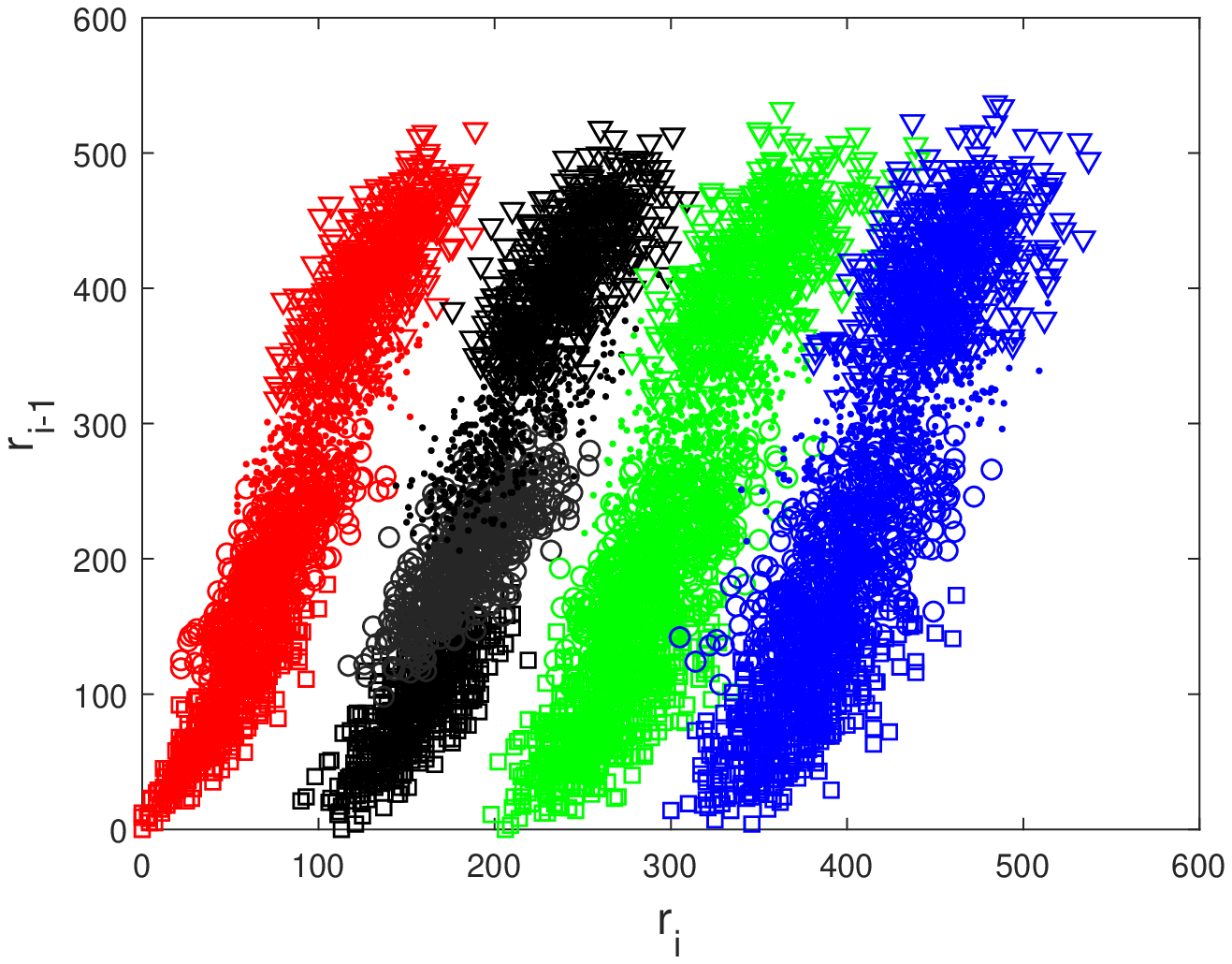}}

\subfigure[ Distribution of \texorpdfstring{$ [r_n,r_{n-1} ]|_{\hat{s}_n,\hat{s}_{n-1}} $}{original} with estimated labels using Algorithm 3]{\label{fig:Kmeans_BER_MultiLevel_3} \includegraphics[width=0.45\textwidth]{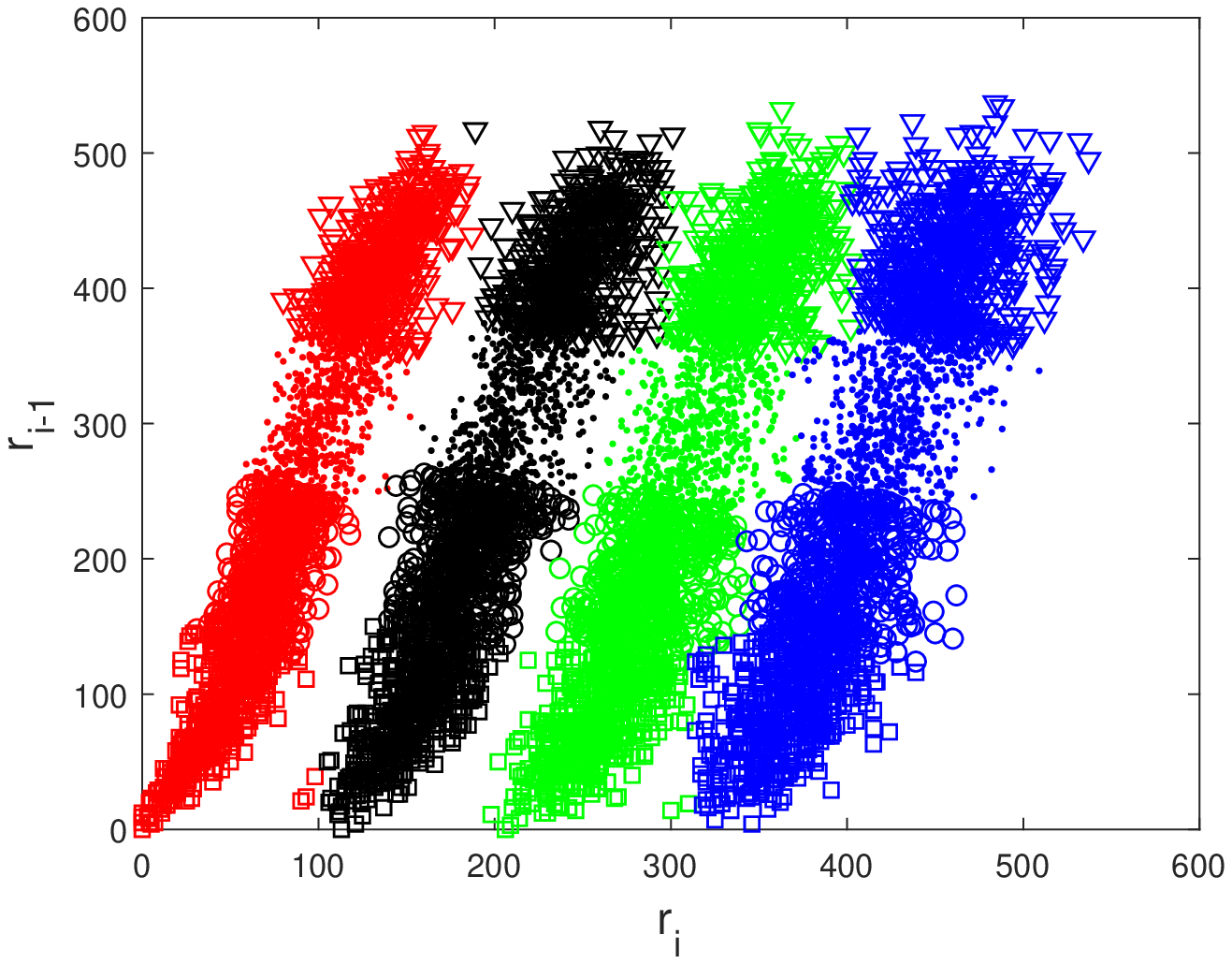}}
\subfigure[ Distribution of \texorpdfstring{$ [r_n,r_{n-1} ]|_{\hat{s}_n,\hat{s}_{n-1}} $}{original} with estimated labels using Algorithm 4]{\label{fig:Kmeans_BER_MultiLevel_4} \includegraphics[width=0.45\textwidth]{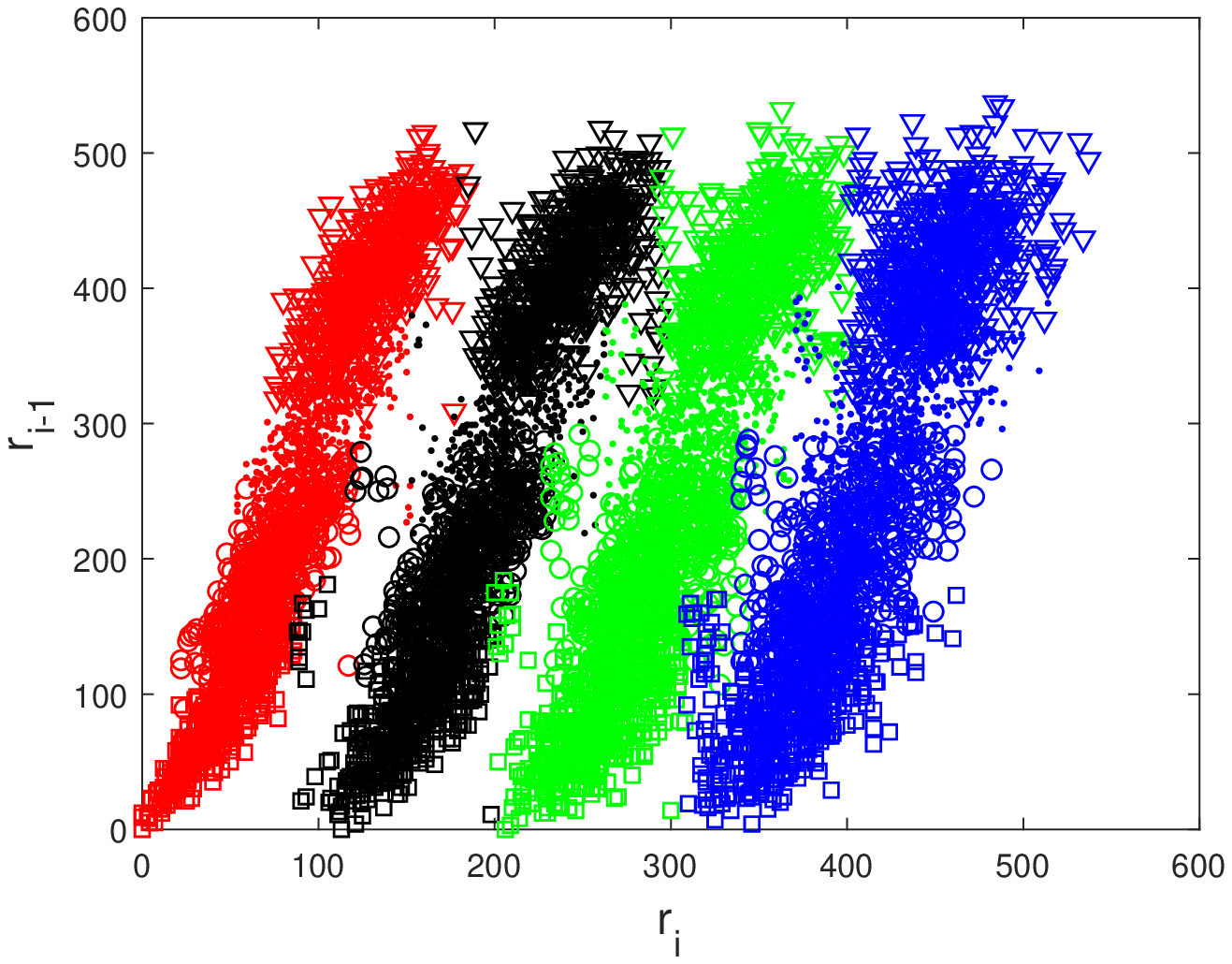}}
\caption{ $ \mathcal{L}=1 $  and $ T = 60\ \Delta T $ }
\label{fig:Kmeans_BER_comparison_MultiLevel}
\end{figure}

In a  4-level CSK modulation scheme, there exist 16 clusters for the points $ [r_n,r_{n-1}] $ as depicted in Fig. \ref{fig:Kmeans_BER_MultiLevel_2}. In this case, we consider Algorithms \ref{alg:iterative_clustering_based_inference} and \ref{alg:iterative_clustering_plus_threshold} for detection, and, in the first iteration, the initial centroids are set to $ \bm{\mu}_0 = 0 $, $ \bm{\mu}_1 = \frac{\max(\bm{r})}{3} $, $ \bm{\mu}_2 = \frac{2\max(\bm{r})}{3} $ and $ \bm{\mu}_3 = \max(\bm{r}) $.

The numerical results are illustrated in Fig. 13. We observe that, in the considered case study, Algorithm 3 outperforms other benchmark algorithms including Algorithm 4. In this case in which multiple symbols are transmitted, in fact, the detection performance of the proposed schemes are inherently more vulnerable to wrong estimations of the decoding thresholds. Therefore, Algorithm 3 outperforms Algorithm 4 in the considered case study. This behavior can be understood by comparing Fig. 13(b) with Fig. 13(c) and Fig. 13(d), in which we evince that Algorithm 3 yields a smaller classification error compared with Algorithm 4.

\section{Conclusion} \label{sec:conclusion}
In this paper, we have introduced non-coherent detection schemes for application to MC systems in the presence of ISI. The proposed algorithms are based on combining clustering methods and empirical estimates of the detection thresholds that are employed in memory-bits detection methods. In order to apply the proposed clustering-based algorithms, different methods for initializing the centroids of the clusters directly from the empirical data are proposed and analyzed. Simulation results show that, in the presence of severe ISI, the proposed iterative algorithms that may combine multi-dimensional clustering methods and detection thresholds yield, in general, good BER performance provided that the a sufficient number of memory bits are used.


\end{document}